%% file: main.tex
\PassOptionsToPackage{prologue,table}{xcolor}
\documentclass[sigconf]{acmart}

\makeatletter
\def\@ACM@checkaffil{% Only warnings
    \if@ACM@instpresent\else
    \ClassWarningNoLine{\@classname}{No institution present for an affiliation}%
    \fi
    \if@ACM@citypresent\else
    \ClassWarningNoLine{\@classname}{No city present for an affiliation}%
    \fi
    \if@ACM@countrypresent\else
        \ClassWarningNoLine{\@classname}{No country present for an affiliation}%
    \fi
}
\makeatother
\AtBeginDocument{%
  \providecommand\BibTeX{{%
    \normalfont B\kern-0.5em{\scshape i\kern-0.25em b}\kern-0.8em\TeX}}}

\acmConference[ICSE 2024]{46th International Conference on Software Engineering}{April 2024}{Lisbon, Portugal}

\setcopyright{none}
\renewcommand\footnotetextcopyrightpermission[1]{} % removes footnote with conference information in first column

\input{macro}

\begin{document}


%\title{What to do next: Context-aware Automated GUI testing via Large Language Model}
% \title{Chatting with GPT-3 for Zero-Shot Human-Like Mobile Automated GUI Testing}

\title{Make LLM a Testing Expert: Bringing Human-like Interaction to Mobile GUI Testing via Functionality-aware Decisions}

% \title{Function-driven Decisions: Bringing Human-like Interaction to Automated Mobile GUI Testing with LLM}
% \title{Bringing Human-like Interaction to Automated Mobile GUI Testing with LLM}
%\title{Empowering Human-Like Mobile App Testing by Chatting with LLM/GPT-3}

\author{Zhe Liu$^{1}$,Chunyang Chen$^2$, Junjie Wang$^{1,*}$, Mengzhuo Chen$^{1}$, Boyu Wu$^{1}$, Xing Che$^{1}$, \\ Dandan Wang$^{1}$, Qing Wang$^{1,*}$}
\affiliation{
  \position{$^1$State Key Laboratory of Intelligent Game, Beijing, China}
  \department{Institute of Software Chinese Academy of Sciences, Beijing, China; \\
  University of Chinese Academy of Sciences, Beijing, China; $^*$Corresponding author\\
  $^2$Monash University, Melbourne, Australia;
  }
}
\email{liuzhe181@mails.ucas.ac.cn, Chunyang.chen@monash.edu, junjie@iscas.ac.cn, wq@iscas.ac.cn}

\begin{abstract}

Automated Graphical User Interface (GUI) testing plays a crucial role in ensuring app quality, especially as mobile applications have become an integral part of our daily lives. 
Despite the growing popularity of learning-based techniques in automated GUI testing due to their ability to generate human-like interactions, they still suffer from several limitations, such as low testing coverage, inadequate generalization capabilities, and heavy reliance on training data. 
Inspired by the success of Large Language Models (LLMs) like ChatGPT in natural language understanding and question answering, we formulate the mobile GUI testing problem as a Q\&A task.
We propose {\tool}, asking LLM to chat with the mobile apps by passing the GUI page information to LLM to elicit testing scripts, and executing them to keep passing the app feedback to LLM, iterating the whole process. 
Within this framework, we have also introduced a functionality-aware memory prompting mechanism that equips the LLM with the ability to retain testing knowledge of the whole process and conduct long-term, functionality-based reasoning to guide exploration. 
We evaluate it on 93 apps from Google Play and demonstrate that it outperforms the best baseline by 32\% in activity coverage, and detects 31\% more bugs at a faster rate.
Moreover, {\tool} identify 53 new bugs on Google Play, of which 35 have been confirmed and fixed.

\end{abstract}

% \begin{IEEEkeywords}
% \end{IEEEkeywords}

\maketitle

\input{sec/introduction}

\input{sec/background}
\input{sec/approach}
\input{sec/effectiveness}

\input{sec/usefulness}
\input{sec/discussion}
\input{sec/related}
\input{sec/conclusion}
\bibliographystyle{ACM-Reference-Format}

\bibliography{reference}

\end{document}

%% file: macro.tex
\usepackage{soul}
\usepackage{float}
\usepackage{threeparttable}
\usepackage{booktabs}
\usepackage{multirow}
\usepackage{epstopdf}
\usepackage{hyperref}
\usepackage{listings}
\usepackage{fancybox}
\usepackage{graphicx}
\usepackage{paralist}
\usepackage{ragged2e}
\usepackage{enumitem}

\usepackage{marvosym}
\usepackage{multirow}
\usepackage{epstopdf}
\usepackage{hyperref}
\usepackage{listings}
\usepackage{fancybox}
\usepackage{graphicx}
\usepackage{subfloat}
\usepackage{cleveref}
\usepackage{paralist}
\usepackage{ragged2e}
\usepackage{enumitem}
\usepackage{pythonhighlight}
\usepackage{algpseudocode}% http://ctan.org/pkg/algorithmicx
\usepackage[normalem]{ulem} % use normalem to protect \emph
\newcommand\hlg{\bgroup\markoverwith
  {\textcolor{green!10}{\rule[-.5ex]{2pt}{2.5ex}}}\ULon}
\newcommand\hlb{\bgroup\markoverwith
  {\textcolor{blue!10}{\rule[-.5ex]{2pt}{2.5ex}}}\ULon}
\newcommand\hlr{\bgroup\markoverwith
  {\textcolor{red!10}{\rule[-.5ex]{2pt}{2.5ex}}}\ULon}

\usepackage[framemethod=TikZ]{mdframed}
\usepackage{tcolorbox}% http://ctan.org/pkg/tcolorbox
\makeatletter
\newcommand{\mybox}[1]{%
  \setbox0=\hbox{#1}%
  \setlength{\@tempdima}{\dimexpr\wd0+13pt}%
  \begin{tcolorbox}[boxrule=0.5pt, colback=white, arc=4pt,
      left=6pt,right=6pt,top=6pt,bottom=6pt,boxsep=0pt]
    #1
  \end{tcolorbox}
}
\usepackage{tikz}
\usepackage{pgfplots}
\usetikzlibrary{pgfplots.statistics,calc}
\usepackage{array}
\usepackage{amsmath}
\usepackage{centernot}
\usepackage{xspace}
\usepackage{url}
\usepackage{verbatim}
\usepackage{wrapfig}
\usepackage{tabularx}
\usepackage{bbding}
\usepackage{pifont}
\usepackage{wasysym}
\usepackage{amssymb}

\newcommand{\tool}{\texttt{GPTDroid}}
\makeatletter  
\newif\if@restonecol  
\makeatother  
\usepackage[linesnumbered,ruled,vlined]{algorithm2e}%[ruled,vlined]{  
\usepackage{algpseudocode}  
\usepackage{amsmath}  

%% file: sec/introduction.tex
\section{Introduction}
\label{sec_introduction}
%Background
In recent years, mobile apps have become an indispensable part of our daily life, with millions of apps available for download from app stores like the Google Play Store~\cite{Googleplay} and Apple App Store~\cite{Appstore}. 
With the rise of app importance in our daily life, it has become increasingly critical for app developers to ensure that their apps are of high quality and perform as expected for users. 
To avoid time-consuming and labor-extensive manual testing, automated GUI (Graphical User Interface) testing is widely used for quality assurance of mobile apps~\cite{mirzaei2016reducing,yang2018static,yang2013grey,machiry2013dynodroid,zeng2016automated,mao2016sapienz}, i.e., dynamically exploring mobile apps by executing different actions such as scrolling and clicking to verify the app functionality. 
% based on the program analysis
%Current solution

%  when testing practical commercial apps
Unfortunately, existing GUI testing tools such as probability-based or model-based ones~\cite{su2021benchmarking,kong2018automated,rubinov2018we} suffer from low testing coverage, meaning that they may miss important bugs and issues. 
This is because of the complex and dynamic nature of modern mobile apps~\cite{kong2018automated,rubinov2018we,rubin2015covert,Rico,pecorelli2022software,fan2018large}, which can have hundreds or even thousands of different screens, each with its own unique set of interactions and possible user actions and logic.
In addition, test inputs generated by these methods are significantly different from real users' interaction traces~\cite{peng2022mubot}, resulting in low testing coverage. 
To address these limitations, there has been a growing interest in using deep learning (DL)~\cite{li2019humanoid,yasin2021droidbotx} and reinforcement learning (RL)~\cite{pan2020reinforcement,romdhana2022deep} techniques for automated mobile GUI testing. 
By learning from human testers' behavior, these methods aim to generate human-like actions and interactions that can be used to test the app's GUI more thoroughly and effectively.
These approaches are based on the idea that the more closely the actions performed by the testing algorithm mimic those of a human user, the more comprehensive and effective the testing will be.

%Limitation of related works
Nevertheless, there are still some limitations with these DL or RL-based GUI testing methods.
First, learning algorithms require large amounts of data which is difficult to collect from real-world users' interactions.
Second, learning algorithms are designed to learn and predict from training data, so they may not generalize well to new, unseen situations, as apps are constantly evolving and updating.
Third, mobile apps can be non-deterministic, meaning that the outcome of an action may not be the same every time it is performed (e.g., clicking the ``delete'' button from a list with the last content would produce an empty list for which the delete button no longer works) which specifically makes it difficult for RL algorithms to learn and make accurate predictions.
Therefore, another more effective approach to generate human-like actions is highly needed to test mobile apps thoroughly.

% \jie{do we need to mention the `human-like' part in the limitations?}

%Basic idea
% The emerging Large Language Model (LLM)~\cite{attention,schulman2022chatgpt,chowdhery2022palm,zhang2022opt} 
% trained on ultra-large-scale corpus, which shows promising performance in natural language understanding, logical reasoning and question answering in recent years. 
% For example, ChatGPT~\cite{schulman2022chatgpt} (Chat Generative Pre-trained Transformer) is one LLM from OpenAI with 175 billion parameters trained on a massive dataset including existing test scripts and bug reports, which makes it capable of understanding and generating text across a wide range of topics and domains. 
% Its success demonstrates that LLM can understand human knowledge and interact with humans as a knowledgeable expert.
% Inspired by ChatGPT, we formulate GUI testing problem as a questions \& answering (Q\&A) task, i.e., asking the LLM to play a role as a human tester to test the target app.

Large Language Models (LLMs)~\cite{attention,schulman2022chatgpt,chowdhery2022palm,zhang2022opt} such as GPT-3/4 have emerged as a powerful tool for natural language understanding and question answering. Recent advances in LLM have triggered various studies examining the use of these models for software development tasks\cite{liu2022fill,kang2023large,xia2022less}
% \chen{\cite{??, ??}}. \chen{May introduce our Qtypist work here, 
% and tell the difference.}
The ChatGPT~\cite{schulman2022chatgpt} (Chat Generative Pre-trained Transformer) from OpenAI, has billions of parameters and is trained on a vast dataset comprising test scripts and bug reports. Its exceptional performance across diverse domains and topics demonstrates the LLM's ability to comprehend human knowledge and interact with humans as a knowledgeable expert. Inspired by ChatGPT, we formulate the GUI testing problem as a questions \& answering (Q\&A) task, i.e., asking the LLM to play the role as a human tester to test the target app.

We propose {\tool} for automated GUI testing, which asks LLM to chat with mobile apps by passing the GUI page information to LLM to elicit testing scripts and execute them to keep passing the app feedback to LLM, iterating the whole process.
To convert the visual information of the app GUI into the corresponding natural language description, we first extract the semantic information of the app and GUI page by decompiling the target app and view hierarchy files, and design linguistic patterns to encode the information as the prompt of LLM.
We then utilize few-shot learning by providing demonstrations with the output template to facilitate the LLM generating desired executive commands to execute the app.
% \rev{given the natural language described answer from LLM, we decode it into actionable steps to execute the target app.} 

Nevertheless, there are two main challenges during the interactive Q\&A GUI testing process. 
The first is the local dilemma. 
Different from the LLM-based program repair or unit test generation which mainly targets a determined piece of software, {\tool} formulates the GUI testing as a multi-turn task and the LLM faces varying GUI pages, i.e., interacting between LLM and mobile app to explore various pages of the app.
During the interaction process, it is hard for the LLM to clearly and accurately remember the historical explorations, especially those that happened long before. 
Because of this, the LLM might only rely on the recent interactive information to make the decision, while omitting the global viewpoints, which can make the exploration fall into a local dilemma and hinder it from achieving higher coverage. 
The second is the low-level dilemma.
Being fed with the descriptive GUI information, the LLM can easily focus more on the low-level semantics as the widgets or activities, yet less on the high-level semantics as the functionalities which is achieved with sequences of operations with the widgets/activities.  
However, the functional aspect of the mobile app is of high interest to testers and users, and has long been an obstacle to existing techniques.

To overcome these challenges, within {\tool}, we develop a functionality-aware memory mechanism. 
It builds a testing sequence memorizer to record all the interactive testing information in terms of the explored activities and widgets. 
It also queries the LLM about the function-level progress of the testing during the iterative process, e.g., which function is under test, to enable the LLM to conduct the explicit reasoning by itself. 
And the information is then encoded into a functionality-aware memory prompt and fed into the LLM to enable the LLM in deciding the meaningful operation sequence to explore the app's functionality and conduct global exploration to cover unexplored areas.

\input{figure/example-of-prompt.tex}

%benefit
% \rev{Compared with conventional learning-based algorithms mentioned above, our approach is based on LLM as a zero-shot tester that does not require any training data or corresponding computational resources for training the model.}
% \jie{the above  statement is a little questionable, i think maybe we can delete it.}
One example chat log can be seen in Figure~\ref{fig:Example-prompt}.
LLM can understand the app GUI, and provide detailed actions to navigate the app (e.g., A1-A5 at Figure~\ref{fig:Example-prompt}).
To compensate for its wrong prediction (A2 at Figure~\ref{fig:Example-prompt}), the real-time feedback by {\tool} guides it to regenerate the input until triggering a valid page transition.
It remains clear testing logic even after a long testing trace to make complex reasoning of actions (A3, A4 at Figure~\ref{fig:Example-prompt}), and it can prioritize to test important functions earlier (e.g., A5 at Figure~\ref{fig:Example-prompt}). 
A more detailed analysis of {\tool}'s capability is in Section~\ref{Sec_Discussion}.

To evaluate the effectiveness of {\tool}, we carry out an experiment on 93 popular Android apps in Google Play with 143 bugs.
Compared with 10 common-used and state-of-the-art baselines, {\tool} can achieve more than 32\% boost in activity coverage and 20\% boost in code coverage than the best baseline, resulting in 75\% activity coverage and 66\% code coverage. 
As {\tool} can cover more activities, the method can detect 31\% more bugs with a faster speed than the best baseline.
% \chen{Need to update once Section 4.3.3 is settled}.
%Through further analysis, we find three findings and advantages for improving the performance of the method: (1) For text input, {\tool} can generate text with practical meaning. (2) {\tool} can reasonably generate combined operations. (3) {\tool} can be targeted to test important functions.
Apart from the accuracy of our {\tool}, we also evaluate the usefulness of our {\tool} by detecting unseen crash bugs in real-world apps from Google Play. 
Among 223 apps, we obtain 53 crash bugs with 35 of them being confirmed and fixed by developers, while the remaining are still pending. 
To reveal reasons behind the promising performance of our approach, we further investigate the experiment results qualitatively and summarize 4 findings including function-aware exploration through long meaningful testing trace,function-aware prioritization, valid text input and compound action.
% \jie{need revise here after the discussion section is done.}

The contributions of this paper are as follows: 
% \chen{I make it short, one sentence for each point.}
\begin{itemize}

\item \textbf{Vision.} 
The first work to formulate the automatic GUI testing problem to an interactive question \& answering task to let the LLM conduct the whole app testing by understanding the GUI semantic information and automatically inferring possible operation steps.
% The first work to formulate the automated GUI testing problem to a question \& answering task by bringing LLM into GUI testing domain, to the best of our knowledge.%to assist GUI testing tools in achieving higher testing coverage. %It opens a new dimension for automated GUI testing by directly using LLMs as testing strategy generation engines.

\item \textbf{Technique.} 
A function-aware automatic GUI testing approach {\tool} which designs the function-aware memory mechanism to enable the LLM to focus more on the global and functional viewpoints of the mobile app. 
% A novel approach {\tool}\footnote{We release the source code, dataset, and detailed experimental results on our website \textit{https://github.com/testinging6/GPTDroid}. And  We further provide a demo video \url{https://youtu.be/5hnLKqxZ904} to facilitate the understanding of {\tool}. \label{github}} based on ``pre-train, prompt and predict'' paradigm of the LLM by understanding the GUI semantic information and dynamic context of the iterative testing process, for automatically inferring possible operation steps.

\item \textbf{Evaluation.} Effectiveness and usefulness evaluation of {\tool} in the real-world apps with practical bugs detected (Section \ref{sec_Effectiveness} and \ref{sec_Usefulness}).%with promising results. 

\item \textbf{Insight.} Detailed qualitative analysis revealing the reasons why LLM can generate human-like and functionality-aware actions for app testing (Section \ref{Sec_Discussion}). %Effectiveness evaluation of {\tool} with high activity coverage. We further uncover the finding of the essential reasons for promising performance, including realizing complex combination operations and exploring deeper pages. The usefulness evaluation of {\tool} with 25 confirmed and fixed GitHub issues\footnote{We release the source code, dataset, and detailed experimental results in our website \textit{https://github.com/Socrates/Socrates} to facilitate the replication and follow-up studies. \label{github}}. \chen{Tell more about insights/findings, while less about the experiment.}

\end{itemize}

%% file: figure/example-of-prompt.tex
\begin{figure}[t]
\centering
\vspace{0.05in}
\includegraphics[width=8.4cm]{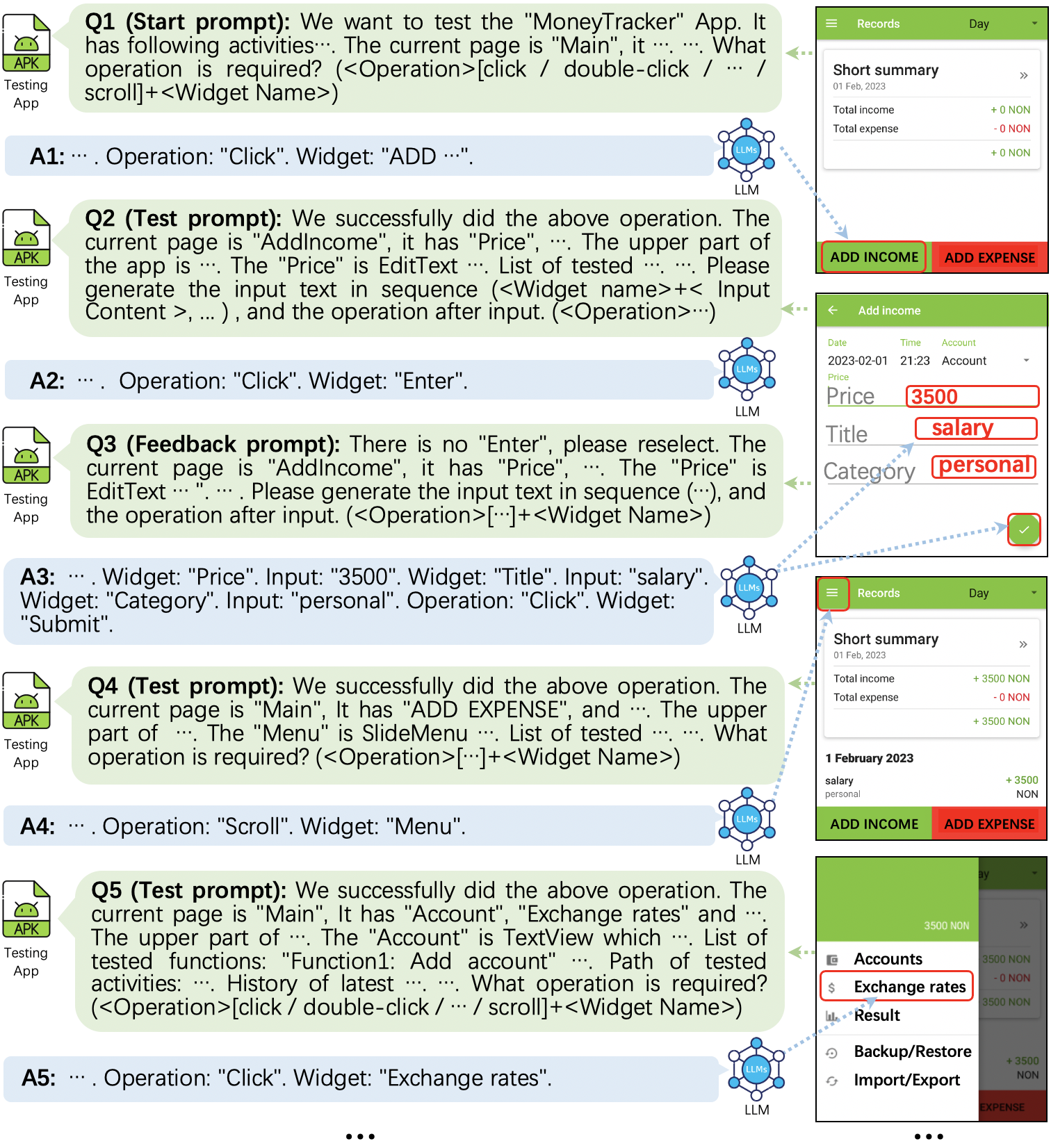}
\vspace{-0.1in}
\caption{Demonstrated example of how {\tool} works.}
% \jie{comment}}
%这里answer里面，function也不要了吧，感觉刚开始这块也不太能读懂。就直接 省略号，operation就行了。
%然后把问题里面 function相关的也不要了。都用省略号。如果嫌内容少，描述页面的可以多一点。就是这个图就只涉及外层循环相关的就行，让大家知道怎么work的，其他的就省略号了。
%这是我个人的意见哈，具体你自己看着来。我觉得在这个时候，大家get不到太多细节，有太多细节反而容易confusing。这个图主要是要大家看到总体的一些逻辑。
% \chen{Can we use PDF file to ensure its visibility.}}
% \jie{comment}
%A4里面的or，感觉好像解决不了了吧。只说用and这些分割。我觉得去掉比较好。
%这个A5里面，accounts本来就是第一个，感觉其他工具也是首先测，所以可能有其他例子吗？
%这个图，我觉得不能全部都是往图上面指的箭头，应该图里面的指到prompt，表示抽取信息生成prompt，这个可以用蓝色；然后answer指到图上面做操作，用红色，这样感觉比较好 是吧？
%再就是这些文字看得挺清晰的，图感觉稍微有点不清晰啊。能不能把文字这块横向再缩小一些，纵向会潜在拉长。然后图这块也放大一点点。可能会稍微好一点？
% }
\label{fig:Example-prompt}
\vspace{-0.15in}
\end{figure}

%% file: sec/approach.tex
\section{Approach}
\label{sec_approach}

We model the GUI testing as a Question \& Answering (Q\&A) problem, i.e., asking the LLM to play a role as a human tester, and enabling the interactions between the LLM and the app under testing.
To realize this, we propose {\tool}, as demonstrated in Figure \ref{fig:overview}, with nested loops.

In the outer loop, it extracts the GUI context information of the current GUI page, encodes them into prompt questions for LLM, decodes LLM's feedback answer into actionable operation scripts to execute the app, and iterates the whole process.
Specifically, in each iteration of the testing, {\tool} first obtains the view hierarchy file of the mobile app and extracts the GUI context including the app information, information of the current GUI page, and details of each widget in the page. 
We then design linguistic patterns for generating the GUI prompt as input of LLM. 
We utilize the idea of few-shot learning to enable the LLM's output to conform with our expected standards which can be directly executed in the app, by providing the demonstrations as reference.

In the inner loop, it builds a testing sequence memorizer to record all the detailed interactive testing information, e.g., the explored activities and widgets. 
During the process, the memorizer also stores the functionality-level progress of testing, e.g., which function is under test, which is derived by querying the LLM and is to enable LLM to conduct the explicit reasoning by itself. 
We also design linguistic patterns to encode the information into the functionality-aware memory prompt, to equip LLM with the capability of retaining knowledge of the whole testing and conducting the long-term reasoning. 
The prompt in both the outer loop and inner loop would together input into the LLM for querying the next operation.

% \liuzhe{It extracts the GUI context information of the current GUI page and constructs the functional chain of thought in testing process, encodes them into prompt questions for LLM, decodes LLM's feedback answer into actionable operation scripts to execute the app; and iterates the whole process. }
% % It extracts the static and dynamic context of the current GUI page, encodes them into prompt questions for LLM, decodes LLM's feedback answer into actionable operation scripts to execute the app; and iterates the whole process. 
% Armed with the knowledge learned from large-scale training corpus, {\tool} would have the potential to guide the testing in exploring more diversified pages, conducting more complex operational actions, and covering more meaningful operational sequences.  

\input{figure/overview.tex}

\input{tab/App-GUI-information.tex}

\subsection{GUI Context Extraction}
\label{subsec_approach_information_Extraction}

Despite its excellence on various tasks, the performance of LLM can be significantly influenced by the quality of its input, i.e., whether the input can precisely describe what to ask~\cite{chen2022knowprompt,liao2022ptau,zhou2022learning}.
In the scenario of this interactive mobile GUI testing, we need to accurately depict the GUI page currently under test, as well as its contained widgets information from a more micro perspective, and the app information from a more macro perspective.
% \rev{Furthermore, to act like a human tester, {\tool} should also capture current testing progress so as to recommend the testing operations from a more global viewpoint to potentially cover more activities and avoid duplicate explorations. }
This Section describes which information will be extracted, and Section \ref{subsec_approach_Prompt_Generation} will describe how we organize the information into the style that LLM can better understand. 
GUI context relates to the information of the app, the GUI page currently tested, and all the widgets on the page. 
The app information is extracted from the \textit{AndroidMaincast.xml} file, while the other two types of information are extracted from the view hierarchy file, which can be obtained by UIAutomator~\cite{uiautomator}.
Table \ref{tab:app-GUI-info} presents the summarized view of them. 

\textbf{App information} provides the macro-level semantics of the app under testing, which facilitates the LLM to gain a general perspective about the functions of the app. 
The extracted information includes the name of the app and the name of all its activities.

\textbf{Page GUI information} provides the semantics of the current page under testing during the interactive process, which facilitates the LLM to capture the current snapshot.  
We extract the activity name of the page,  all the widgets represented by the ``text'' field or ``resource-id'' field (the first non-empty one in order), and the widget position of the page.
For the position, inspired by the screen reader~\cite{zhang2021screen,stangl2021going,wang2021screen2words}, we first obtain the coordinates of each widget in order from top to bottom and from left to right, and the widgets whose ordinate is below the middle of the page is marked as lower, and the rest is marked as upper.

\textbf{Widget information} denotes the micro-level semantics of the GUI page, i.e., the inherent meaning of all its widgets, which facilitates the LLM in providing actionable operational steps related to these widgets. 
The extracted information includes ``text'', ``hint-text'', and ``resource-id'' field (the first non-empty one in order), ``class'' field, and ``clickable'' field.
To avoid the empty textual fields of a widget, we also extract the information from nearby widgets to provide a more thorough perspective, which includes the ``text'' of parent node widgets and sibling node widgets.
% \jie{comment}
%我觉得觉得你新加的那句话不太需要。因为你后面不是已经说了 我们从nearby widgets区信息了。所以我删了。

% \subsubsection{\textbf{Dynamic Context Extraction}}
% \label{subsubsec_approach_Information_Extraction}

% Dynamic context relates to the detailed testing progress, which facilitates the LLM being well aware of the process context and making informed decisions.  
% We design an \textbf{operation memorizer} \liuzhe{in Section \ref{subsec_approach_Output-Matching}} to keep the record of this information, \liuzhe{including Long-term memory and Short-term memory, i.e., the test path of activity, the functions that have been tested, the summary of GUI pages tested in the last five steps, and the number of operations, as shown in Table \ref{tab:app-GUI-info}.}

% Specifically, during the iteration, when an operation is conducted, we can obtain the widget information of the operation, and the GUI pages information after the operation, and then the operation memorizer is updated accordingly. 
% In detail, the visit number of the widget is updated by finding the same widget in the operation memorizer by the ``text'' field and ``resource-id'' field of the widget.
% The visit number of the GUI page is updated by finding the same activity in the memorizer with the ``ActivityName'' field of the page.
% % }

\input{tab/approach-rule.tex}

\subsection{\textbf{GUI Prompting and Executive Command Generation}}
\label{subsec_approach_Prompt_Generation}

% \jie{comment}
%感觉有两个generation不太好，我把prompt generation改成了prompting。后面那个sec我也修改了。
% \chen{``Operation Matching'' sounds a bit weird, what about executive command generation?}
With the extracted information, we design linguistic patterns to generate prompts for inputting into the LLM. 
We first conduct preprocessing for the information, to facilitate the follow-up design. We tokenize attributes by the underscore and Camel Case~\cite{pascalcase} considering the naming convention in app development.
% We then conduct the part-of-speech (POS) tagging with Stanford NLP parser~\cite{de2008stanford}, and only retain the noun, verb and prepositions for the linguistic patterns. 

\subsubsection{\textbf{Linguistic Patterns of GUI Prompt}}
\label{subsubsection_patterns_of_prompt}
To design the patterns, each of the five annotators is asked to write the prompt sentence following regular prompt template~\cite{chen2022knowprompt,Cantino201Prompt,gu2021ppt}, and questions the LLM for generating the operation steps. 
We then check to what extent the recommended operation is reasonable considering the whole testing process.
% Each annotator can access a random-chosen 100 apps from Google Play, and they can obtain the preprocessed static context information and the dynamic context information. 
% After 10 hours of trial, we are required to provide the most promising and diversified 20 prompt sentences, which are served as the seeds for designing patterns. 
With the prompt sentences, the five annotators then conduct card sorting~\cite{spencer2009card} and discussion to derive the linguistic patterns.
As shown in Table \ref{tab:approach-rule}, this process comes out with 6 linguistic patterns corresponding with the three sub-types of information in Table \ref{tab:app-GUI-info} and two operation \& feedback patterns.

\textbf{Pattern related to GUI context (Table \ref{tab:approach-rule}-Id 1,2,3}) We design three patterns to describe the overview of the GUI page currently under testing, respectively corresponding to the app information, page GUI information, and widget information in Table \ref{tab:app-GUI-info}.

% \textbf{Pattern related to dynamic context:} We design one pattern to describe the testing progress with the dynamic context as shown in Table \ref{tab:app-GUI-info}. 

\textbf{Pattern related to operation \& feedback question (Table \ref{tab:approach-rule}-Id 4,5,6})  
% \jie{comment}
%我觉得这一句感觉不是pattern，这个不涉及到prompt的pattern吧，这个不是输出吗？我觉得就放到operation matching那里就行。
We also design patterns to describe operation and feedback questions. For the operational questions, we ask the LLM what operation is required. 
We also provide the output template in the prompt to enable the LLM to generate a desired executive command for testing the app, and details are in Section \ref{subsec_approach_Output-Matching}.
Note that, we separate the patterns for querying general action (e.g., click) and text input (e.g., input certain text) with \textit{pattern-Id 4 and 5} respectively, to enable the generated commands can better match the widgets in the GUI page. 
For the feedback question, after deciding the previous operation is not applicable,
% (as described in Section \ref{subsec_approach_Output-Matching}), \jie{comment}
%现在这个sec 应该没有这个信息。
we inform the LLM that there is no such widget on the current page, and let it re-try. 
% \liuzhe{And for the memorize question, we ask the LLM during each interaction what is the current tested function and whether to start testing a new function. If LLM answers yes, we will record the tested function names(as described in Section \ref{subsec_approach_Output-Matching}). If LLM answers no, it indicates that the previous function is still being tested.}

\subsubsection{\textbf{Prompt Generation Rules}}
\label{subsubsection_Prompt_generation_rules}

Since the designed patterns describe information from different points of view, we combine the patterns from different viewpoints and generate the prompt rules as shown in Table \ref{tab:approach-rule}.
We design three kinds of prompts respectively for starting the test, routine inquiry, and getting feedback in case of error occurred.
Note that, due to the robustness of the LLM, the prompt sentence doesn't need to follow the grammar completely.

\textbf{Test prompt} is the most commonly used prompt for informing the LLM of the current status and query for the next operation. 
Specifically, we 
% \rev{first tell the LLM the dynamic context, i.e., how many times each GUI page and widget has been explored; followed by} 
tell the LLM the GUI context, i.e., the information about the current GUI page and detailed widget information; then ask the LLM which operation is required.

\textbf{Feedback prompt} is used for informing the LLM error occurred and re-try for querying the next operation.
Specifically, we first tell LLM its generation operation cannot correspond to the widget on the page; re-provide it the detailed widget information of the page and let the LLM recommend the operation again.

Besides the above two kinds of prompts, we additionally design \textbf{start prompt} to start the testing of the app and only used it once. 
Different from the test prompt, it provides the LLM with the app information including all activities for a global overview.
% \rev{But it does not have any dynamic context since the testing just begins.}
% \rev{This prompt is only used once. since the LLM can somehow remember this global app information during the testing process}
% ~\cite{li2019humanoid,su2017guided}. 

\input{tab/CoT-prompt}
\subsubsection{\textbf{Executive Command Generation}}
\label{subsec_approach_Output-Matching}

% \chen{Suggest to refer to Sidong's ICSE'24 work for this section. Do you need to tell ChatGPT what actions do you support? And ask ChatGPT to select corresponding actions accordingly.}

After inputting the generated prompt, LLM will output the natural language sentence of operation, e.g., \textit{input 3500 for price, input salary in title widget and personal in category widget, then click submit which depicts the example output for the second image in Figure \ref{fig:Example-prompt}}. 
Considering that a testing operation can be expressed in different ways and with different words, it is challenging to map natural language testing operations to the app for execution. 
Therefore, we utilize the idea of in-context learning to provide LLM with the output template, including available operations and operation primitives, which can be mapped directly to the instructions for executing the app. 
% We also employ the few-shot learning by providing demonstrations to the LLM to enable it better understand the output template.  
% \rev{They can fix the output format of LLM for further mapping to components and operations.}

% \jie{comment}
%你这块不是已经限定了输出格式了，那就不是nlp described sentence了。
%就是这段到后面的转折比较突兀，不清楚为啥你要定义available operations，以及primitives。你这一段就要说明白，你这个整体逻辑是啥，所以你需要定义opeartions和primitives
%感觉这里应该先说对于LLM的输出，面临的最大问题是如何将自然语言描述的输出映射到app上进行执行，为了解决这个问题，我们规范化了输出的格式（或者看看sidong文章是怎么说的），并且采用了in context learning，通过让大模型看到一些输出格式的例子，从而让他产生复合格式的输出，得到的输出可以直接去app执行。

\textbf{Available operations.} We identify five commonly used standard operations for mobile applications, including click, double-click, long press, scroll and input. Although there are other customizations, such as custom gestures, they are not common, and we leave them for future work.
% . For the sake of brevity, we will focus on the commonly used operations in this paper.

\textbf{Output templates.}
We design different operation primitives for the above-mentioned operations to represent widgets and executive commands.
% \jie{comment}
%represent entities 这个说法 略微突兀。不知道为啥有个entities
The above five operations can be divided into two main categories, i.e., \textit{action} (the first four operations) and \textit{input}. 
% Specifically, for the action category, it mainly covers clicking, scrolling, double clicking, and long pressing. 
For the action category, we formulate it as <Operation>[click / double-click / long press / scroll] + <Widget name>, e.g., \textit{Operation: ``Click''. Widget: ``ADD INCOME''.}. 
% The scrolling can have up, down, left or right options, while <Widget Name> represents ``up'', ``down'', ``left'' and ``right'' (e.g., Operation: ``scolling''. Widget:``up''). 
% \jie{comment}
%我们要不把scroll那句话去掉吧，比较细节，怕confusing大家。如果被问到再说。
% \jie{comment}
% %这个不应该是个name吗，为啥还有up and down，是说name后面加着up 或者 down？还是怎么着
% \liuzhe{comment}
% % 滑动一般是页面的滑动，不需要组建名，直接确定滑动的方向就行。所以这里组建名就是Up/down
For the input category, the GUI page usually involves text field widgets for entering specific values, and the follow-up operations (usually for submitting the text input).
We formulated it as <Widget name> + <Input content>, e.g., \textit{Widget: ``Price'', Input: ``3500''}, followed by <Operation> + <Widget name>, e.g., Operation: ``Click''. Widget: ``Submit'' .
Table \ref{tab:prompt works} provides an example answer from LLM for these two categories.

\subsection{Functionality-aware Memory Prompting}
\label{subsec-chain-of-thought}

% \chen{I have read this part thoroughly, but cannot get the idea. 1)Is this CoT? It seems that we just collect some data and organize them into a nice way. So it should be counted as part of prompt engineering in the last subsection. 2)Since we have uploaded the first version to Arxiv, so please re-run all experiments to reflect the changes.}
% \liuzhe{comment}
% 1) 感觉也算是一种chain of thought，我们不断迭代更新我们测试路径，作为prompt来实时告诉LLM
% 2）目前的实验结果和我们Arxiv上的已经不同，实验的数据集也进行了更新

With the context extraction, GUI prompting and executive command generation in the previous two sections, {\tool} can already conduct the automated GUI testing. This Section proposes the function-aware memory prompting, which further improves the capability of the approach in retaining the knowledge during the iterative testing process and understanding the functional aspects of the mobile app, so as to generate the function-aware operations to guide the operation from global viewpoints. 

To achieve this, we build a testing sequence memorizer (Section \ref{Testing Sequence Memorizer}) to record all detailed testing information.
We also query the LLM about the function-level progress of the testing (Section \ref{Function-level Progress}) in each iterative testing step, and store it into the memorizer.
We then design linguistic patterns to encode the information into the functionality-aware memory prompt (Section \ref{Long-term Chain of Thought Prompt}), and query the LLM together with the prompt in the previous section. 

% \rev{Considering that mobile apps have rich functions and each function has a complete business logic chain. During the process of testing an app, testers often focus on function and determine the next steps based on the testing path and function-level progress. Inspired by this, we use the chain of thought approach to help LLM understand the function-level progress (what function is currently being tested), testing path and latest testing history, to make reasonable judgments.}

\subsubsection{\textbf{Functionality-level Progress}}
\label{Function-level Progress}

Due to the importance of functionalities for app users, we hope our automated GUI testing can conduct the exploration from the viewpoints of the functionalities.
Hence we need to make LLM understand what function is currently being tested derived from the explored activities and widgets.
Specifically, we design a prompt to ask LLM what functionality is currently being tested and whether it has been completed in Section \ref{Long-term Chain of Thought Prompt}. 
And the LLM would provide us with the functionality-level progress as <Function name> + <Status>, in which `YES' denotes it has completed testing the specific functionality. 

To facilitate the LLM to make reasonable functionality-level decisions, we also provide the LLM with the list of functionalities of the app in the prompt. 
This information is first extracted from the app description file and the activity names to serve as the initial seed, and will continuously let the LLM output the refined information during the iterative testing process. 

% Firstly, LLM output the core functions of the app as an ``inspiration'' (global information) based on its app description and the activity names before the testing. Please note that these functions may be incomplete and inaccurate, and LLM will continuously iterate and update the function-level progress in real-time during the subsequent testing process. Then, during the testing process, we design prompt to ask LLM what functionality is currently being tested and whether it has been completed in Section \ref{Long-term Chain of Thought Prompt}. Finally, LLM provide us the function-level progress result as <Function name> + <Status>.

\subsubsection{\textbf{Testing Sequence Memorizer}}
\label{Testing Sequence Memorizer}

We design a testing sequence memorizer to keep the record of testing, including the set of tested functions (in Section \ref{Function-level Progress}), the testing path of activity, the set of tested activities with page visits number, the set of tested widgets of the current page with widgets visits number.

Specifically, during the iteration, when an operation is conducted, we can obtain the status of the function (<Function name>+<Status>), the operation of the widget (<Operation>+<Widget name>), and the test path of activity (<Activity1>+<Operation>+<Activity2>)). Then the operation memorizer is updated accordingly. 
In detail, the visit number of the widget is updated by finding the same widget in the operation memorizer with the ``text'' field and ``resource-id'' field of the widget.
The visit number of activity is updated by finding the same activity in the memorizer with the ``ActivityName'' field.
% of the page.

% \subsubsection{\textbf{Function-level progress}}
% \liuzhe{Due to the importance of functions in apps, we need to make LLM understand what function is currently being tested and its testing status (i.e., whether the functionality has been tested). Specifically, we designed corresponding questions in the prompt to ask LLM during each testing process. Firstly, we will extract the core functions of the application based on its description information. Then input these functions into LLM as global information for the app. Finally, during the testing process, we asked LLM what functionality is currently being tested and whether it has been completed. LLM will provide us with the current functionality under testing and the completion status of the testing based on the issues in the prompt.}
% \jie{comment}
% %we will extract the core functions of the application based on its description information. 这个是怎么做的？我觉得这个就挺难的，一般app对于自己functions的描述也都不太全，基本上也就列出非常少的几个功能，那你是这块是怎么做的？只能依赖于app文档列出的功能吗？LLM能自己基于activity name这些总结出来一些function吗
% \liuzhe{comment}
% % 这块我做的时候实际上是让LLM首先根据description和activity name生成App可能的功能（因为描述中一般会写App的主要功能），然后这些功能作为一个参考，加入到我们chain of thought中。我写的时候在想这样会不会太复杂了，就改成我们按照关键词提取直接来获取可能的功能，因为只做一次所以效果是一样的，用LLM的就是在图上和描述起来可能有点复杂。

\subsubsection{\textbf{Functionality-aware Memory Prompt}}
\label{Long-term Chain of Thought Prompt}

Based on the information in the testing sequence memorizer, we further construct a functionality-aware memory prompt to facilitate the LLM to keep an eye on the functionality during recommending the next operation. 
It consists of three parts, including the explored functionalities, the covered activities, and the recently tested operations, and it would refer to the testing sequence memorizer in each iteration to fetch the latest information.  
We follow the same procedure in Section \ref{subsec_approach_Prompt_Generation} to derive these prompt patterns. 
The details and examples of the prompts are shown in Table \ref{tab:memory prompt}.
% The above information is updated after each iteration, and LLM will make targeted judgments based on the long-term chain of thought.

\textbf{Pattern related to explored functionalities (Table \ref{tab:memory prompt}-Id 1).}
This describes which functions have been explored, the number of explorations, and whether it is finished in testing the function.
Since the GUI prompt in the previous Section only demonstrates the widgets and activities in the current GUI page, it could not provide the high-level viewpoints of the functionality which is accomplished by a sequence of operations on the widgets/activities.
Therefore, this prompt can remind the LLM about the functional aspects of the app, and facilitate the LLM in deciding the meaningful operation sequence to explore the app's functionality. 
Specifically, {\tool} extracts the tested functions from the testing sequence memorizer, including the visit times of function pages, and the testing status of the functions.
% we extract the function name from the output of LLM, and record the number of times the function was explored. 
% \jie{comment}
%最后这句不太对。这里应该是在memorizer已经存储了，就不是直接从output of LLM取了，而且output of LLM只是单次的。所以你这里得说memorizer记录了 status of functions，然后你是怎么得到这个信息的。这块我觉得细节不对。你再memorizer 是不是要对function name进行合并，怎么判断 number of times the function was explored。
%好像是status也是从llm output 获取的，但其他就不是了。这里你要梳理下到底memorizer存储的是啥。这里prompt又是做了什么操作。

\textbf{Pattern related to covered activities. (Table \ref{tab:memory prompt}-Id 2)}
This describes the sequence of covered activities during the testing process.
Since the basic component of the mobile app is the activity which is recorded in the \textit{AndroidManifest.xml} file, this prompt aims at providing the activity viewpoints of testing history to enable the LLM better capture the tested functionalities and to cover more unexplored areas.
Specifically, we merge the adjacent same activity and activity sequence, and update the number of visits accordingly.

\textbf{Pattern related to recently tested operations. (Table \ref{tab:memory prompt}-Id 3)}
This denotes the latest test page with the detailed visiting status of all its contained widgets, as well as the operations leaving the page. 
This information is the first-hand and fine-grained testing recording, which can facilitate the LLM in capturing the latest status of the testing progress, and make informed decisions about exploring a certain functionality. 
% Specifically, each time GPTDroid executes a test script, we record the widgets that have been visited on each page and the number of times the widgets have been visited. 
Specifically, {\tool} extracts the widgets tested on each GUI page and their visit times from the testing sequence memorizer.
% \jie{comment}
%这里我们需要说record 信息这个事情吗，感觉是不是应该在memorizer那里，就已经说明吧了，我们每次操作之后都会record了。这里我觉得只需要说，基于memorizer我们做了些什么操作，来组装成了prompt
For the current test page, we will select the operation pages of the lastest \textit{k} steps and the corresponding operations. For each page, we provide LLM with the activity name of the current page and the number of visits to each widget when the page is accessed.
We set \textit{k} as 5 based on the empirical experience.

% \jie{comment}
%按照前面那个章节的逻辑，是不是还得有个 Patterns related to functionality？ Function question patterns。
%那个叫做 functionality inquiry pattern吧
%到最后时候，尽量把这些pattern在表格的编号 也在正文引用一下。
%而且这个prompt，我看你也有output template，那也是用的in-context learning吗？是怎么能让他生成规定的格式的。这里是不是也要提一下，例如就是说我们采用和前面相同的in-context learning来得到desired output。

\textbf{Pattern related to functionality inquiry: (Table \ref{tab:memory prompt}-Id 4)} 
We ask the LLM what function is currently tested, and also provide the output template in the prompt to enable the LLM to generate the function name and its status, with details in Section \ref{Function-level Progress}.

% \jie{comment}
%上面这段我不细看了，反正主要concern是他们和memorizer的关系，以及memorizer到底存的什么内容，什么格式。这块你自己把握下。

\textbf{Prompt Generation Rules.} 
We combine the above three patterns for providing LLM with functionality aspects of testing information from different viewpoints, and generate the prompt rules as shown in Table \ref{tab:memory prompt}. We provide examples of how the GUI context prompts and memory prompts work to enable app testing.

\input{tab/prompt-example}

\subsection{Implementation}
\label{subsec_approach_Implement}
% \chen{Can you please write the implementation details following the order of the approach introduction, also mention which section to help readers understand.}

{\tool} is implemented as a fully automated GUI app testing tool, which uses or extends the following tools: VirtualBox~\cite{virtualbox} and the Python library pyvbox~\cite{pyvbox} for running and controlling the Android-x86 OS, Android UIAutomator~\cite{uiautomator} for extracting the view hierarchy file, and Android Debug Bridge (ADB)~\cite{Adb} for interacting with the app under test (Section \ref{subsec_approach_information_Extraction}).
% \liuzhe{commet}
% 这里可能要讨论一下
For the LLM (Section \ref{subsec_approach_Prompt_Generation}), we use 
% \rev{We implement our testing large language model based on}
the pre-trained ChatGPT model which was released on the OpenAI website\footnote{\url{https://platform.openai.com/docs/models/gpt-3-5}}.  
The basic model of ChatGPT is the \textit{gpt-3.5-turbo} model which is extremely powerful and good at answering questions.

%% file: figure/overview.tex
\begin{figure}[!th]
\centering
\vspace{-0.05in}
\includegraphics[width=8.3cm]{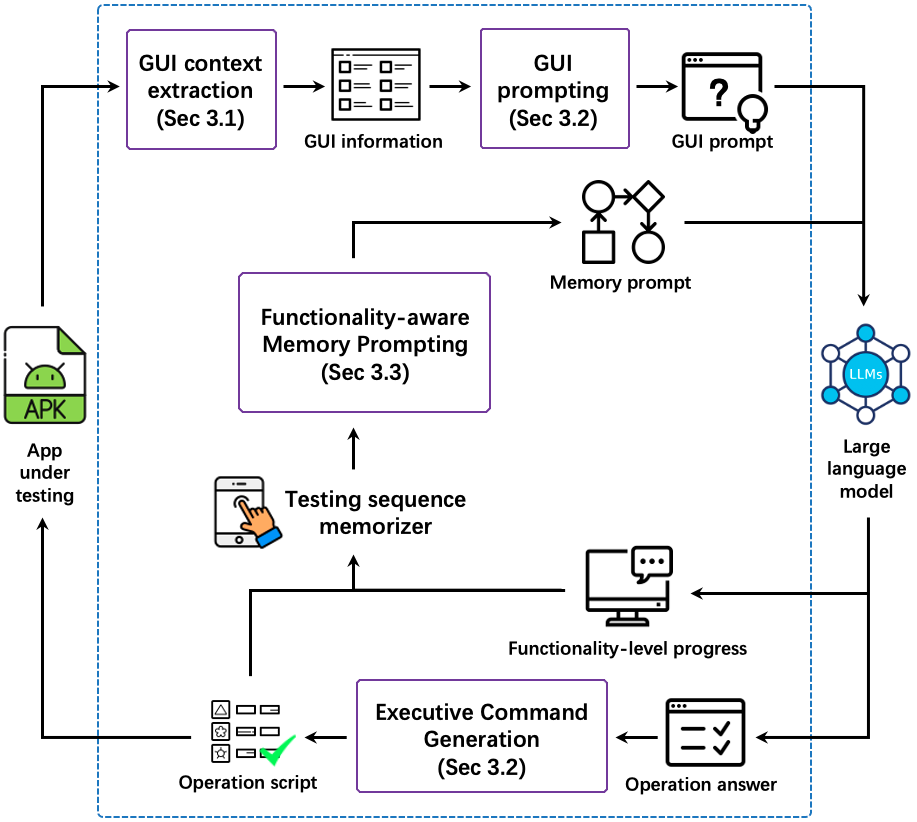}
\vspace{-0.1in}
\caption{Overview of {\tool}.}
% \chen{I suggest to not add ``ChatGPT'' in this figure.}}
% \jie{comment}
%这个图上内容上次春阳老师不是说要简化下吗，static context extraction (Sec 3.1) 那个static可以用斜体，后面dynamic也是, dynamic context extraction 
%testing operation memorizer，要不把testing去掉，这个字号可以大一点，或者怎么突出下？这个和其他的例如operation script，其实意义还是不太一样。
%那个record，要不要换成update？
%gui prompt 我们要换成gui prompt question吗？下面那个feedback answer，这样正好对应下Q&A
% }
% \jie{comment+}
%图里面framework of our approach删掉。
%改成‘static context extration’‘dynamic context extraction’
%感觉这里面很多字号都可以再增大一些，例如线上的文字，还有现在的Q1-Q4，A1-A4里面的内容，好像也看不太清楚，是否也加大个字号？
% }
% \jie{comment}
%这个图基本是沿用了input generation那个工作的逻辑，但其实这两个工作是有明显区别的。那个input generation就是针对一个页面，怎么建模，然后输入内容，就完了。而这个工作是一个复杂任务，就是需要对整个app进行测试，他是和大模型有频繁交互的。类似于张俊林知乎里面提到的复杂任务，需要分而治之这种。
%而且其实感觉现有工作中，很多都还是在解决单一的小任务，包括春阳老师发的Enabling Conversational Interaction with Mobile UI 那个工作，他也就是单次和LLM交互，我们这个是个复杂任务，所以具体怎么表达这个事情，还要好好想想。
%你这里的error correction mechanism其实是一个方面，但我感觉好像也就是 please select again体现了 error correction吧。
%你提到的context 这个事情，你说的widget context 我其实觉得比较片面。那个input generation的工作，因为我们是要为某个input输入内容，所以可以说建模这个input附近的context。但这里，我觉得更应该说是整个测试的context，就是现在已经执行到什么阶段了，有个state tracking这个感觉的意思。现在这个图感觉还是在说单步的逻辑，我觉得这个图更应该体现总体的逻辑。
% } 
\label{fig:overview}
\vspace{-0.15in}
\end{figure}

%% file: tab/App-GUI-information.tex
\begin{table*}
\vspace{-0.05in}
\renewcommand\arraystretch{0.95} 
\caption{Extracted GUI information and examples.}
% \chen{1. For ID 5, layout, or should be called "position". 2. For ID 7, will the text of the textview will also be considered? What if it is very long?}\liuzhe{comment}}
% 这里我们采用了全部的，目前textview的长度对prompt影响不是特别大，没有遇到过超出的情况
\vspace{-0.1in}
\label{tab:app-GUI-info}
\centering
% \footnotesize
\scriptsize
\begin{center}
\begin{tabular}{m{0.2cm}<{\centering} | m{1.3cm}<{} | m{6.8cm}<{} | m{7.7cm}<{}}
\toprule
\textbf{Id} & \textbf{Attribute} & \textbf{Description}  & \textbf{Examples} \\
\midrule
\multicolumn{4}{c}{\textbf{GUI context - App information}} \\
% \rowcolor{red!10}
\midrule
1 & AppName & Name of the app under testing & AppName = ``Money Tracker'' \\
2 & Activities & List of names for all activities of the app, obtained from \textit{AndroidManifest.xml} file & Activities = [``Main'', ``AddAccount'', ``Import'', ``Income'', ...] \\
\midrule
\multicolumn{4}{c}{\textbf{GUI context - page GUI information}}\\ 
% \rowcolor{blue!10}
\midrule
3 & ActivityName & Activity name of the current GUI page & ActivityName = ``AddPersonalInformation'' \\
4 & Widgets & List of all widgets in current page, represented with text/id & Widgets = [``Edit Account'', ``btn\_income'', ...] \\
5 & Position & Relative position of widgets, obtained through their coordinates & Upper = [``Welcome'', ...], Lower = [``Add Income'', ...] \\
% \jie{confirm : we only use upper and lower? }\\
\midrule
\multicolumn{4}{c}{\textbf{GUI context - widget information}} \\ 
% \rowcolor{green!10}
\midrule
6 & WidgetText & Widget text, obtained by field `text' or `hint-text' & WidgetText = ``Welcome to the Money Tracker!'' \\
7 & WidgetID & Widget ID, obtained by field `resource-id'. & WidgetID = ``add\_account'' \\
8 & WidgetCategory & Category: TextView, EditText, ImageView, etc, obtained by field `class' & WidgetCategory = ``TextView'' \\
9 & WidgetAction & Widget action, obtained by field `clickable', such as click, input, etc. & WidgetAction = ``Click'' \\
10 & NearbyWidget & Nearby widgets, obtained by the text of parent widgets and sibling widgets & NearbyWidget = ``your income: [SEP] \$ '' \\
% \midrule
% \multicolumn{4}{c}{\textbf{Dynamic context - Long-term memory}} \\ 
% % \rowcolor{green!10}
% \midrule
% % 11 & Duplicate & Whether this page is the same as the previous page, obtained through similarity & Duplicate = false \\
% % 11 & PageVisits & Set of tested GUI pages with page visits number & PageVisits = [\{``Main'': ``5'' , ``Account'': ``3'', ``Setting'': ``1'',  ...\}]\\
% % 12 & WidgetVisits & Set of tested widgets of current GUI page with widget visits number & WidgetVisits = [\{``Income'': ``2'', ``Add'': ``2'', ``Delete'': ``3'', ...\}]\\
% 11 & ActVisits & Set of tested activities (GUI pages) with page visits number & ActVisits = [\{``Main'': ``5'' , ``Account'': ``3'',  ...\}]\\
% 12 & ActSeq & Sequence of tested activities (GUI pages) & ActSeq = [``Main'', ``Account'', ``AddAccount'', ``Account'', ...]\\
% 13 & Functions & Set of tested functions & Functions = [``Add Account'', ``Edit Account'', ``Delete Account'', ...]\\
% \midrule
% \multicolumn{4}{c}{\textbf{Dynamic context - Short-term memory}} \\ 
% \midrule
% 14 & WidgetVisits & Set of tested widgets of current GUI page with widget visits number & WidgetVisits = [\{``Income'': ``2'', ``Add'': ``2'', ...\}]\\
% 15 & 5PagesSeq & Sequence of 5 tested GUI pages & 5PagesSeq = [\{``5-page'': [\{``Activity'':``Main'', VisitWidgets:''Account, ...''\}], ``4-page'': [...] ,  ..., ``1-page'': [...] ,\}]\\
% 16 & 5Operations & Set of tested operations of each activities & 5Operation = [\{``[5]Add account'': click, ``[4]Income'': [``input1'':``300''], ..., ``[1]Delete'': click,\}]\\

\bottomrule
\end{tabular}
\end{center}
\vspace{-0.05in}
\end{table*}

%% file: tab/approach-rule.tex
\begin{table*}
% \vspace{-0.1in}
\renewcommand\arraystretch{0.95} 
\caption{The example of linguistic patterns of GUI prompts and generation rules.}
% \jie{comment}
%我觉得要不要在pattern那里，或者notes那里，说一下，如果gui page上面有editText widget，就用input operation；否则就用action operation。我觉得你括号那个input，是没啥意义，而且在例子那里也没有了。
%type那里，我觉得id 4 就叫做 querying general action。id 5 叫做 querying text input； 
%后面那个table 3里面，id 4叫做 functionality inquiry
% }
\vspace{-0.1in}
\label{tab:approach-rule}
\centering
\footnotesize
% \scriptsize
% \begin{center}
\begin{tabular}{p{0.2cm}<{\centering} | p{2.0cm} | p{6.3cm}<{} | p{7.9cm}<{}}
\toprule
\textbf{Id} & \textbf{Pattern type} & \textbf{Sample of linguistic patterns/rules} & \textbf{Instantiation}\\
\midrule
\multicolumn{4}{c}{\textbf{GUI context patterns: \hlr{\textit{GUIContext}}}}\\ 
% \rowcolor{red!10}
\midrule
\rowcolor{red!10}
1 & App information & We want to test the <\textit{AppName}> App. It has the following activities, including <\textit{Activities}>. & We want to test ``Money tracker'' App. It has the following activities, including ``Main'', ``AddAccount'', ``Import'', ``Setting'', ... . \\ 
\rowcolor{red!10}
2 & Page GUI information & The current page is <\textit{ActivityName}>, it has <\textit{Widgets}>. The upper part of the app is <\textit{Position}>, the lower part is <\textit{Position}>. & The current page is ``Main'', it has ``Income'', ``Add'', ``Delete'',  ... . The upper part of the app is ``Welcome to ..., Delete, ...'', the lower part of the app is ``Income, ...''.  \\ 
\rowcolor{red!10}
3 & Widget information & The widgets which can be operated are <\textit{WidgetText / WidgetID}>. <\textit{WidgetText/WidgetID}> is <\textit{WidgetCategory}> which can <\textit{WidgetAction}> and its nearby widget is <\textit{NearbyWidget}>.  & The widgets which can be operated are ``Add'', ``Delete'', ``Edit Account'',... . ``Add'' is Button which can be clicked and its nearby widget is  ``Add account, ...'' , ``Delete'' is TextView which can be clicked and its nearby widget is ... . \\ 
\midrule
\multicolumn{4}{c}{\textbf{Operation \& feedback question patterns: \hlg{\textit{OperationQuestion}}}} \\ 
\midrule
 \rowcolor{green!10}
4 & Querying general action & Action operation question + <\textit{Output Template}>& What operation is required? (<\textit{Operation}>[click / double-click / long press / scroll]+<\textit{Widget Name}>) \\
 \rowcolor{green!10}
5 & Querying text input & Input operation question + <\textit{Output Template}>& Please generate the input text in sequence, and the operation after input. (<\textit{Widget name}>+<\textit{Input Content}>, ...) and provided (<\textit{Operation}[click]>+<\textit{Widget name}>) \\
 \rowcolor{green!10}
6 & Testing feedback & Feedback question & There is no <\textit{WidgetText / WidgetID}> on the current page, please reselect. \\ 
\bottomrule
\toprule
\multicolumn{4}{c}{\textbf{Prompt generation rules}}\\
\midrule
1 & \multicolumn{3}{l}{\textbf{Start Prompt:} \hlr{\textit{GUIContext}[1,2,3]} + \hlg{\textit{OperationQuestion}[4/5]}}\\
\midrule
2 & \multicolumn{3}{l}{\textbf{Test Prompt:} We successfully did the above operation. \hlr{\textit{GUIContext}[2,3]} + \hlg{\textit{OperationQuestion}[4/5]}}  \\
\midrule
3 & \multicolumn{3}{l}{\textbf{Feedback Prompt:} Sorry,  \hlg{<\textit{OperationQuestion}>[6]} + \hlr{<\textit{GUIContext}>[3]} + \hlg{<\textit{OperationQuestion}>[4/5]} } \\
\bottomrule
\end{tabular}
\begin{tablenotes}
% \footnotesize
\scriptsize
\item \textbf{\textit{Notes:}} ``[1,2, ..., 6]'' means the id of each pattern. If there is ``EditText'' on the page, GPTDroid selects ``Querying text input'' pattern. For the other widgets, GPTDroid selects ``Querying general action'' pattern.
\end{tablenotes}
% \end{center}
\vspace{-0.1in}
% \vspace{-0.15in}
\end{table*}

%% file: tab/CoT-prompt.tex
\begin{table*}
\vspace{-0.05in}
\renewcommand\arraystretch{0.95} 
\caption{ The example of linguistic patterns of functionality-aware memory prompts and generation rules.}
\vspace{-0.1in}
\label{tab:memory prompt}
\centering
\footnotesize
% \scriptsize
\begin{center}
\begin{tabular}{p{0.2cm}<{\centering} | p{1.8cm}<{} | p{6.6cm}<{} | p{7.3cm}<{}}
\toprule
\textbf{Id} & \textbf{Type} & \textbf{Sample of linguistic patterns/rules} & \textbf{Instantiation} \\
\midrule
\multicolumn{4}{c}{\textbf{Long-term functionality-aware memory patterns:\hlr{$FunctionMemory$}}} \\ 
% \rowcolor{red!10}
\midrule
\rowcolor{red!10}
% 1 & Initial possible functions of app (possible priority order): <Function Name\langle$, ... & Initial possible functions of app (possible priority order): ``Add your income'', ``Add Account'' ... \\
% \rowcolor{red!10}
1 & Explored functionalities & \textbf{List of tested functions:} <\textit{Function Name}> + <\textit{Visits Time}> + <\textit{Status}>, <\textit{Function Name}> + <\textit{Visits Time}> + <\textit{Status}> ... & \textbf{List of tested functions:} ``Function1: Add your income. Visits: 3. Status: Finished'', ``Function2: Delete information. Visits: 2. Status: Finished'', ... \\
\rowcolor{red!10}
2 & Covered activities. & \textbf{Path of tested activities:} <\textit{Activity Name}> + <\textit{Visits Time}>, <\textit{Activity Name}> + <\textit{Visits Time}>, ...  & \textbf{Path of tested activities:} ``Activity: Main.  Visits: 3'', ``Activity: Account.  Visits: 4'',  ``Activity: AddAccount.  Visits: 3'', ...  \\
\rowcolor{red!10}
3 & Recently tested operations & \textbf{History of latest tested pages and operations:} \textbf{Latest 5th step} tested the <\textit{Activity Name}> page. The following widgets with visits time of this page have been tested: <\textit{Widget Name}> + <\textit{Visits Time}>, ... . The following executive command achieve the page transition: <\textit{Operation}> + <\textit{Widget Name}>. & \textbf{History of latest tested pages and operations:} \textbf{Latest 5th step} tested the ``Exchange'' page. The following widgets with visits time of this page have been tested: ``Widget: Add exchange, Visits:2'', ``Widget: Submit exchange, Visits:1'', `Widget: Cancel, Visits:1'' ... . The following executive command achieve the page transition: ``Click'' the "Exchange".  \\
\rowcolor{red!10}
& & \textbf{Latest 4th step} tested the <\textit{Activity Name}> page. The following ... & \textbf{Latest 4th step} tested the ``main'' page. The following widgets ... \\
\rowcolor{red!10}
& &... & ... \\
\rowcolor{red!10}
& & \textbf{Latest 1st step} tested the <\textit{Activity Name}> page. The following ... & \textbf{Latest 1st step} tested the ``Add Account'' page. The following widgets ... \\
\midrule
\multicolumn{4}{c}{\textbf{Function question patterns: \hlg{$FunctionQuestion$}}} \\ 
\midrule
 \rowcolor{green!10}
4 & Functionality inquiry & Functionality inquiry + <\textit{Output Template}> & What is the functions currently being tested? Are we testing a new function? (<\textit{FunctionName}>  + <\textit{Status}>)  \\ 
\bottomrule
\toprule
\multicolumn{4}{c}{\textbf{Functionality-aware memory prompt generation rules}}\\
\midrule
1 & \multicolumn{3}{l}{\textbf{Functionality-aware memory Prompt:} \hlr{\textit{FunctionMemory}[1,2,3]} + \hlg{\textit{FunctionQuestion}[4]}} \\

\bottomrule
\end{tabular}
% \begin{tablenotes}
% \footnotesize
% \item \textbf{\textit{Notes:}} ``[1,2, ..., 5]'' means the id of each pattern. 
% \end{tablenotes}
\end{center}
\vspace{-0.1in}
% \vspace{-0.15in}
\end{table*}

%% file: tab/prompt-example.tex
\begin{table}[htb]
\vspace{0.05in}
\caption{The example of how prompts work in {\tool}.}
\vspace{-0.05in}
\label{tab:prompt works}
\centering
\footnotesize
\begin{tabular}{p{1.6cm}<{} | p{6.2cm}<{}}
\toprule
\textbf{Prompt type} & \textbf{Instantiation} \\ 
\midrule
\multicolumn{2}{c}{\textbf{Example 1: For general action operation}}\\ 
\midrule
GUI context & \textbf{Start Prompt:}[\hlr{We want to test “Money tracker” App, It has the following activities, including ... .]} \textbf{\textcolor{blue}{or}} \textbf{Test Prompt:} [We successfully did the above operation.] \textbf{\textcolor{blue}{or}} \textbf{Feedback prompt:} \hlg{[There is no ``Exchange'' on the current page, please reselect.]} \\
& \hlr{[The current page is ``AddAccount'', it has ... . The upper part of the app is ... , the lower part ... The ``Exchange'' is Button ...]}  \\
\midrule
Functionality-aware memory & \hlr{[\textbf{List of tested functions:} ``Function1: Add your income. ...'', ...] [\textbf{Path of tested activities:} ``Activity: Main.  Visits: 3'', ...  ] [\textbf{History of latest tested pages and operations:} \textbf{Latest 5th step} tested the ``Exchange'' page. The following widgets ...  \textbf{Latest 1st step} tested the ``AddAccount'' page. ...]}  \\
\midrule
Function question & \hlg{What is the functions currently being tested? Are we testing a new function? (<Function name> + <Status>)} \\
\midrule
Action question & \hlg{What operation is required? (<\textit{Operation}>[click / double-click / long press / scroll]+<\textit{Widget Name}>)} \\
\midrule
\textbf{LLM Answer} & Function: "Add income". Status: Yes. \\
& Operation: "Click". Widget: "ADD INCOME". \\ 
\midrule
\midrule
\multicolumn{2}{c}{\textbf{Example 2: For text input operation}}\\ 
\midrule
GUI context & \textbf{Start Prompt:}[\hlr{...}] \textbf{\textcolor{blue}{or}} \textbf{Test Prompt:}. \textbf{\textcolor{blue}{or}} \textbf{Feedback prompt:} ..\\
\midrule
functionality-aware memory & \hlr{[\textbf{List of tested functions:} ``Function1: Add ...'', ...] [\textbf{Path of} \textbf{tested activities:} ...] [\textbf{History of latest tested pages ...}}] \\
\midrule
Function question & \hlg{What is the functions currently being tested? Are we testing a new function? (<Function name>  + <Status>)} \\
\midrule
Input question & \hlg{Please generate the input text in sequence, and the operation after input. (<\textit{Widget name}>+ <\textit{Input Content}>, ... and provided <\textit{Operation}>+ <\textit{Widget name}>) } \\
\midrule
\textbf{LLM Answer} & Function: "Add income". Status: No. \\
& Widget: "Price". Input: "3500". Widget: "Title". Input: "salary". Widget: "Category". Input: "personal". Operation: "Click". Widget: "Submit". \\
\bottomrule
\end{tabular}
\vspace{-0.15in}
\end{table}

%% file: sec/effectiveness.tex
\section{Effectiveness Evaluation}
\label{sec_Effectiveness}
In order to verify the performance of {\tool}, we evaluate it by investigating the activity and code coverage (RQ1), as well as the number of detected bugs (RQ2).
We also present the ablation study of each module in {\tool} (RQ3).
Note that, this Section utilizes the previously-detected bugs in the app's repositories to demonstrate the effectiveness of {\tool}, and the next Section will evaluate the usefulness of {\tool} in detecting new bugs. 

\subsection{Experimental Setup}
\label{subsec_experiment_dataset}
The experimental dataset comes from two sources. 
The first is from the apps in the Themis benchmark ~\cite{su2021benchmarking}, which contains 20 open-source apps with 34 bugs in GitHub. 
Considering the small number of apps in the benchmark, we collect a second dataset following similar procedures as the benchmark.
% \rev{the second newer and larger dataset, including recently updated and popular commercial apps from Google Play, to further verify the performance of our approach. }
% \jie{comment}
%因为你之前不是还用了写app 进行数据扩充吗，而且不是还有些app让作者产生prompt。我觉得这里要不要提一下，对于满足以上1-3,5 这几个条件，但不满足4,6的（最好把4和5交换下顺序，就是满足1-4，不满足5-6的），总共有多好个app，我们用于在前面的prompt产生，以及训练数据扩充。这样整个逻辑可能比较顺？要不的话，你前面只是说了那些数据没有做实验，那他们是怎么选的，感觉就没有一个标准。

\input{tab/dataset.tex}

\input{tab/RQ1-result.tex}

In detail, we crawl the 50 most popular apps of each category from Google Play~\cite{Googleplay}, and we keep the ones with at least one update after May. 2022, resulting in 407 apps in 12 Google Play categories. 
Then, we use 10 commonly used and state-of-the-art automated GUI testing tools (details are in Section \ref{subsec_experiment_baseline}) to run these apps, in turn, to ensure that they work properly. We then filter out the unusable apps by the following criteria: 
% (1) UIAutomator~\cite{uiautomator} can not obtain the view hierarchy file. 
(1) They would constantly crash on the emulator. (2) One or more tools can not run on them. 
(3) The registration and login functions cannot be skipped with scripts~\cite{su2021benchmarking,lv2022fastbot2,dong2020time}. 
(4) They do not have issue records or pull requests on GitHub. 
% \chen{Why (4)?}
% \liuzhe{comment}
%这里是为了确定bug都是在GitHub上有被确认的bug

There are 73 apps (with 109 bugs) remaining for this effectiveness evaluation.
Note that, same as the benchmark, all bugs are crash bugs.
Specifically, for each app, we select the version in which the bugs are confirmed by developers (merged GitHub pull requests) as our experimental data, following the practice of the benchmark.
% \rev{Finally, 66 apps with 93 bugs remain for further experiments. }
The details of all 93 experimental apps (20 + 73) and related bugs are shown in Table \ref{tab:app-info}.

% \input{tab/RQ1-result.tex}

% \chen{Still not easy to understand the dataset collection in different sections. Can you create a flow-chart to show the process and numbers? It can replace Table 3.}
Note that, there are 101 apps that are filtered out for effectiveness evaluation, yet can successfully run with our proposed approach. 
We apply them to the manual prompt generation in Section \ref{subsubsection_patterns_of_prompt}. 
% and heuristic training data generation in Section~\ref{subsubsec-Heuristic-based Training Data Generation}.
And this ensures that there is no overlap between the apps in approach design and evaluation.

We employ activity coverage, code coverage and the number of detected bugs, which are widely used metrics for evaluating GUI testing~\cite{he2020textexerciser,liu2017automatic,arnatovich2018mobolic,wang2020combodroid,wang2021vet,li2019humanoid}.
We also present the number of covered activities and widgets which are also commonly-used metrics in Table \ref{tab:RQ1-result}. 
% \chen{We rarely use ``;'' in English, so please try to avoid it in our writing.}
% \chen{We may not need to mention code/branch coverage, since I do not have very good ideas. But can you please explain in more details about activity coverage and widget coverage?}
% Following existing studies~\cite{pan2020reinforcement,Android,su2017guided}, 
We treat the activities defined in the \textit{AndroidManifest.xml} file of an Android app as the whole set of  activities~\cite{pan2020reinforcement,Android,su2017guided}. 
During the testing process, we collect the unique activity name and widget ID of the GUI page with which the operation interacts, and treat them as the activity number and widget number. 
% \jie{comment}
%这块这些参考文献，你最后的时候，再核实一下，确保引用对了。那个工作确实是用了这样的指标。要不再被认为fool the reviewers

\subsection{Baselines}
\label{subsec_experiment_baseline}

% \jie{have slightly refined this sec} 

To demonstrate the advantage of {\tool}, we compare it with 10 common-used and state-of-the-art automated testing techniques. 
We roughly divide them into random-/rule-based, model-based, and learning-based methods, to facilitate understanding.
For random-/rule-based methods, we use Monkey~\cite{Monkey} and Droidbot~\cite{li2017droidbot}, Time-Machine~\cite{dong2020time}. For model-based methods, we use WCTester\cite{zheng2017automated,zeng2016automated}, Stoat~\cite{su2017guided}, Ape~\cite{gu2019practical}, Fastbot~\cite{cai2020fastbot}, ComboDroid~\cite{wang2020combodroid}. For learning-based methods, we use Humanoid~\cite{li2019humanoid} and Q-testing~\cite{pan2020reinforcement}. 
 
For a more thorough comparison, we additionally include 5 other baselines, in which the originally proposed techniques aim at enhancing the automated GUI testing, and can be utilized by integrating with the above-mentioned automated testing tools.
Specifically, QTypist~\cite{liu2022fill} is used to generate valid text input to enhance the coverage of automated testing tools. Toller~\cite{wang2021infrastructure} is a tool consisting of infrastructure enhancements to the Android operating system. 
Vet~\cite{wang2021infrastructure} is used to identify exploration tarpits by recognizing their patterns in the UI traces, so as to optimize the exploration sequences.
We use the experimental setup as their original paper to derive the following 5 baselines, i.e, QTypist is integrated with Droidbot and Ape (Droidbot+QT, Ape+QT), Toller is integrated with Stoat (Stoat+TO), Vet is integrated with WCTester and Ape (WCTester+VE, Ape+VE).

We deploy the baselines and our approach on a 64-bit Ubuntu 18.04 machine (64 cores, AMD CPU) and evaluate them on Google Android 7.1 emulators. Each emulator is configured with 2GB RAM, 1GB SDCard, 1GB internal storage, and X86 ABI image. Different types of external files (including PNGs / MP3s / PDFs / TXTs / DOCXs) are stored on the SDCard to facilitate file access from apps. 
Following common practice~\cite{gu2019practical,li2017droidbot}, we registered separate accounts for each bug that requires login and wrote the login scripts, and during testing reset the account data before each run to avoid possible interference. In order to ensure fair and reasonable use of resources, we set up the running time of each tool in one app to 60 minutes, which is widely used in other GUI testing studies~\cite{su2021benchmarking,fan2018large,li2017droidbot,gu2019practical}.
We run each tool three times and obtain the highest performance to mitigate potential bias.

\subsection{Results and Analysis}
\label{subsec_results}
\subsubsection{\textbf{Performance of Coverage (RQ1)}}
\label{sec_results_RQ1}

% \jie{comment}
%code coverage 没提。到时候排版时候，把table 4放到figure 3前面吧，这样也比较符合整体出现的顺序。
Table \ref{tab:RQ1-result} shows the number of covered widgets, number of covered activities, and average activity coverage of {\tool} and the baselines. 
We can see that {\tool} covers far more widgets and activities than the baselines, and the average activity coverage achieves 75\% and average code coverage achieves 66\% across the 93 apps. 
It is 32\% (0.75 vs. 0.57) activity coverage higher even compared with the best baseline (Ape with QTypist). Meanwhile, on the Themis benchmark and Google Play datasets, it was 28\% (0.69 vs. 0.54) and 33\% (0.77 vs. 0.58) higher than the best baseline.
This indicates the effectiveness of {\tool} in covering more activities and codes, thus bringing higher confidence to the app quality and potentially uncovering more bugs. 
Section \ref{Sec_Discussion} will further analyze why {\tool} performs well.

Figure \ref{fig:RQ1-ActivityCoverage} additionally demonstrates the average activity coverage with varying times. 
We can see that, at every time point, {\tool} achieves higher activity coverage than the baselines, and it achieves high coverage within about 24 minutes.
This again indicates the effectiveness and efficiency of {\tool} in covering more activities with less time, which is valuable considering the testing budget. 

\input{figure/RQ1-ActivityCoverage.tex}

Among the baselines, the model-based and learning-based approaches have relatively higher performance. 
Yet the model-based approaches can't capture the GUI semantic information and the exploration could not well understand the inherent business logic of the app.
Learning-based approaches only use little context information for guiding the exploration, and don't have the mechanism for enabling the model considering the app's functionalities.
% \rev{the learners only have limited intelligence restricted by the model architecture and amount of labeled training data. }

We further analyze the potential reasons for the uncovered cases.
% hindering the approach reaching 100\% coverage.  
First, some widgets or inputs do not have meaningful ``text'' or ``resource-id'', which hinders the approach of effectively understanding the GUI page. 
Second, some app requires specific operations, e.g., database connection, long press and drag widgets to a fixed location, which is difficult if not impossible to be automatically achieved.

\input{figure/RQ2-result.tex}

\subsubsection{\textbf{Performance of Bug Detection (RQ2)}}
\label{sec_results_RQ2}

Figure \ref{fig:RQ2-result} shows the overall number of detected bugs of {\tool} and baselines with varying times. 
{\tool} detects 95 bugs for the 93 apps, 31\% (95 vs. 66) higher than the best baseline (Stoat with Toller).
We also compare the similarities and differences of the bugs between Stoat with Toller and our approach, and the results show that all bugs detected by Stoat with Toller are also detected by {\tool}. 
This indicates the effectiveness of {\tool} in detecting bugs and helps to ensure app quality. 

We can also see that, in every time point, {\tool} detects more bugs than the baselines, and reaches the highest value in about 27 minutes, saving 35\% (17 vs. 26) of the testing time compared with the best baseline (also with more detected bugs). 
This again proves the effectiveness and efficiency of {\tool}, which is valuable for saving more time for follow-up bug fixing.
We will conduct a further discussion about the reason behind the superior performance in Section \ref{sec_results_RQ3}.

% \rev{Compared with the baselines, {\tool} can cover more activities \jie{ with valid text inputs and compound operations, and also can guide explore long and meaningful operation sequence, }which enables it in uncovering more bugs. 
% We will conduct further discussion in Section}
% \ref{sec_results_RQ3}.

\input{tab/RQ3-1-result}

\subsubsection{\textbf{Ablation Study (RQ3)}}
\label{sec_results_RQ3}

\textbf{Contribution of Modules.} 
Table \ref{tab:RQ3-1-compents} shows the performance of {\tool} and its 2 variants. 
In detail, for \textit{{\tool} w/o GUI Context} (Sec \ref{subsec_approach_information_Extraction}), we replace the GUI context information with the raw view hierarchy file and extracted the widgets name.
For \textit{{\tool} w/o Function Memory} (Sec \ref{subsec-chain-of-thought}), we remove the functionality-aware memory prompting.
% \jie{comment}
%这里w/o后面的内容要和表格里面完全一致，我已经修改了，大小写你后面可以再统一一下。

We can see that {\tool}'s activity and code coverage are much higher than all other variants, indicating the necessity of the designed modules and the advantage of our approach.
Compared with {\tool}, \textit{{\tool} w/o GUI Context} results in the largest performance decline, i.e., 77\% drop (0.17 vs. 0.75) in activity coverage.
This further indicates that the GUI context extraction can help LLM understand the structure and semantic information of GUI pages and make reasonable judgments.
\textit{{\tool} w/o Function Memory} also undergoes a big performance decrease, i.e., 55\% (0.34 vs. 0.75) in activity coverage. 
This implies our proposed functionality-aware memory prompt can help retain the knowledge during the testing process and gain global viewpoints to reach the uncovered areas. 
% This may be because without the functionality-aware memory, LLM cannot know what functions were performed in the previous step and which pages have been tested.

\textbf{Contribution of Sub-modules.}
Table \ref{fig:RQ3-2-result} further demonstrates the performance of {\tool} and its 8 variants. 
We remove the part of prompt when querying LLM, i.e., the first four variants respectively remove \textit{pattern 1, 2, 3, 5, 6} of Table \ref{tab:approach-rule}, and the last three variants respectively remove \textit{pattern 1, 2, 3} of Table \ref{tab:memory prompt}. 
% \jie{comment}
%这里表格里面的名称，稍微调整下，和table里面的说法大概类似。然后顺序调整下，按我上面说的顺序。
%我上面引用的pattern id 好像不是特别对，你再调整下。总之按照先table 2，再table 3的顺序，每个table里面按照从上到下的顺序。
% We remove each sub-module of the {\tool} in Figure \ref{fig:overview} separately, i.e., App information, page GUI information, widget information, feedback prompt, input operation question, explored functionalities, covered activities and recent tested operations.. For other variants, we set the removed content as ``null''.

% \input{tab/RQ3-2-result}
\input{figure/RQ3-2-result}

% \jie{comment}
%这里有个问题是，为啥module那里的指标是number和activity covery，到了sub module 就变成了activity coverage和code coverage。我觉得要统一一下。
The experimental results demonstrate that removing any of the sub-modules would result in a noticeable performance decline, indicating the necessity and effectiveness of the designed sub-modules. 
Removing the explored functionalities (\textit{{\tool} w/o-Explored Function}) has the greatest impact on the performance, reducing the activity coverage by 51\% (0.37 vs. 0.75). 
This indicates by explicitly querying the LLM about the functionality aspects of the testing progress, the approach can be more aware of what functionality is under test and effectively plan the exploration path to cover more functionalities.

% Provide the explored functionalities for LLM, which can help LLM remember the functions that have been tested.

% \rev{We also notice that, when removing the text input querying prompt (\textit{{\tool} w/o-InputOperation}), the activity coverage also undergoes a large decline by 48\% (0.39 vs. 0.75). 
% The input operation prompt module enables LLM to generate the text input and its operation.}

\input{figure/RQ3-3-result}

\textbf{Influence of Different Number of Recent Tested Operations.}
% \chen{Influence of history length?}
Figure \ref{fig:RQ3-3-result} demonstrates the performance under the different number of latest tested operations. We can see that the activity and code coverage increase with the more tested pages in the latest tested operations, i.e., 5 steps of tested pages and operations.  
And after that, the performance would gradually decrease even increasing the number of tested pages. It also further verified that the 5 steps selected by our pilot study are effective and valuable.
% \jie{comment}
% 这部分删了吧，篇幅超出了太多了。

% \input{tab/RQ3-3-result}

\input{figure/Case-study.tex}

%% file: tab/dataset.tex
\begin{table}[h]
\vspace{-0.05in}
\caption{Dataset of effectiveness evaluation.}
\vspace{-0.05in}
\label{tab:app-info}
\centering
\footnotesize
\begin{tabular}{p{1.2cm}<{\centering} | p{1.2cm}<{\centering}| p{1.2cm}<{\centering} | p{1.2cm}<{\centering} | p{1.2cm}<{\centering}}
\toprule
\textbf{Statistics} & \textbf{\#Activities}  & \textbf{\#Bugs} & \textbf{\#Download} & \textbf{\#Update} \\ 
\midrule
\textbf{Min} & 7 & 1 & 50K+ & 05/22 \\
\textbf{Max} & 33 & 9 & 100M+ & 05/23 \\
\textbf{Median} & 17 & 4 & 5M+ & - \\
\textbf{Average} & 15 & 1.5 & 10M+ & - \\
\midrule
\textbf{All} & 1398 & 143 & - & - \\
\bottomrule
\end{tabular}
\vspace{-0.05in}
\end{table}

%  \begin{table}[h]
% \vspace{-0.05in}
% % \renewcommand\arraystretch{0.95} 
% \caption{Dataset of effectiveness evaluation. \jie{can use this table to present the min, max, median, average, even 1 quarter, 3 quarter of the app number, activity number, XXXX of each app. And in the last row or column, present the total number. This table is uninformative.  }}
% \vspace{-0.05in}
% \label{tab:app-info}
% \centering
% % \footnotesize
% \begin{tabular}{p{4.0cm}<{\centering} | p{3.0cm}<{\centering}}
% \toprule
% &\textbf{The evaluation Apps} \\ 
% \midrule
% \textbf{App number} & 86\\
% \textbf{Activity number} & 690 \\
% \textbf{Update time} & 01/05/2022\\
% \textbf{Downloads} & 1M+ \\
% \textbf{Bug numbers} & 94 \\
% \bottomrule
% \end{tabular}
% \vspace{-0.05in}
% \end{table}

%% file: tab/RQ1-result.tex
\begin{table*}[h]
% \vspace{0.05in}
% \renewcommand\arraystretch{0.95} 
\caption{Performance of activity coverage (RQ1).}
\vspace{-0.05in}
\label{tab:RQ1-result}
\centering
% \footnotesize
\scriptsize
\begin{tabular}{p{2.2cm}<{\centering} || p{0.45cm}<{\centering}  | p{0.45cm}<{\centering}  | p{0.7cm}<{\centering}  | p{0.45cm}<{\centering} || p{0.45cm} <{\centering} | p{0.7cm} <{\centering} | p{0.45cm} <{\centering} | p{0.7cm} <{\centering} | p{0.45cm} <{\centering} | p{0.7cm} <{\centering} | p{0.7cm}<{\centering} | p{0.45cm}<{\centering} || p{0.45cm}<{\centering} || p{0.45cm}<{\centering} | p{0.45cm}<{\centering} || p{0.8cm}<{\centering} }
\toprule
\multirow{2}*{\textbf{Metric}} & \multicolumn{4}{c||}{\textbf{Random-/rule-based}} & \multicolumn{8}{c||}{\textbf{Model-based}} & \multicolumn{3}{c}{\textbf{Learning-based}} \\
 & MK & DB & DB+QT & TM & WC & WC+VE & ST & ST+TO & AP & AP+QT & AP+VE & FB & CD & HM & Q-t & \textbf{{\tool}} \\ 
\midrule
% \multicolumn{4}{c}{\textbf{Random-/rule-based method}}\\
% \midrule
% \textbf{\#Widgets} & 351 & 893 & 1337 & 1582 & 1437 & 1388 & 1701 & 1453 & 1398 & \textbf{2123} \\
% \textbf{\#Activities} & 104 & 269 & 333 & 370 & 391 & 383 & 401 & 340 & 323 & \textbf{517} \\
% \textbf{Avg. activity coverage} & 0.19 & 0.33 & 0.44 & 0.51 & 0.56 & 0.53 & 0.57 & 0.49 & 0.45 & \textbf{0.75} \\
\textbf{\#Widgets}  & 691 & 1707 & 2522 & 3811 & 1745 & 2465 & 2830 & 3087 & 2964 & 3496 & 3406 & 2944 & 3210 & 3022 & 2261 & \textbf{5243}  \\ 
\textbf{\#Activities}  & 266 & 461 & 573 & 811 & 545 & 573 & 615 & 671 & 741 & 853 & 811 & 755 & 783 & 657 & 685 & \textbf{1049} \\ 
\midrule
\textbf{Avg. activity coverage}  & 0.25 & 0.33 & 0.41 & 0.56 & 0.39 & 0.41 & 0.44 & 0.48 & 0.53 & 0.57 & 0.56 & 0.54 & 0.55 & 0.47 & 0.49 & \textbf{0.75} \\ 
\textbf{Avg. code coverage}  & 0.17 & 0.28 & 0.36 & 0.53 & 0.31 & 0.33 & 0.40 & 0.44 & 0.45 & 0.55 & 0.52 & 0.47 & 0.48 & 0.43 & 0.41 & \textbf{0.66} \\ 

% Monkey & 104 & 251 & 0\\
% Droidbot & 269 & 690 & 33\\
% \midrule
% \multicolumn{4}{c}{\textbf{Model-based method}}\\
% \midrule
% Stoat & 333 & 737 & 41 \\
% Ape & 370 & 882 & 39 \\
% Fastbot & 391 & 831 & 47 \\
% ComboDroid & 383 & 783 & 43 \\
% TimeMachine & 401 & 901 & 67 \\
% \midrule
% \multicolumn{4}{c}{\textbf{Learning-based method}}\\
% \midrule
% Humanoid & 340 & 758 & 38 \\
% Q-testing & 323 & 698 & 32 \\
% \midrule
% \textbf{{\tool}} & \textbf{477} & \textbf{989} & \textbf{132} \\
\bottomrule
\end{tabular}
% \vspace{-0.1in}
\begin{tablenotes}
% \footnotesize
\scriptsize
\item \textbf{\textit{Notes:}} ``MK'' is Monkey, ``DB'' is Droidbot, ``DB+QT'' is Droidbot with QTypist, ``TM'' is TimeMachine, ``WC'' is WCTester, ``WC+VE'' is WCTester with Vet,  ``ST'' is Stoat, ``ST+TO'' is Stoat with Toller, ``AP'' is Ape, ``AP+QT'' is Ape with QTypist, ``AP+VE'' is Ape with Vet, ``FB'' is Fastbot, ``CD'' is ComboDroid, ``HM'' is Humanoid, ``Q-t'' is Q-testing. 
\end{tablenotes}
\end{table*}

%% file: figure/RQ1-ActivityCoverage.tex
\begin{figure}[htb]
\centering
\vspace{-0.05in}
\includegraphics[width=8.2cm]{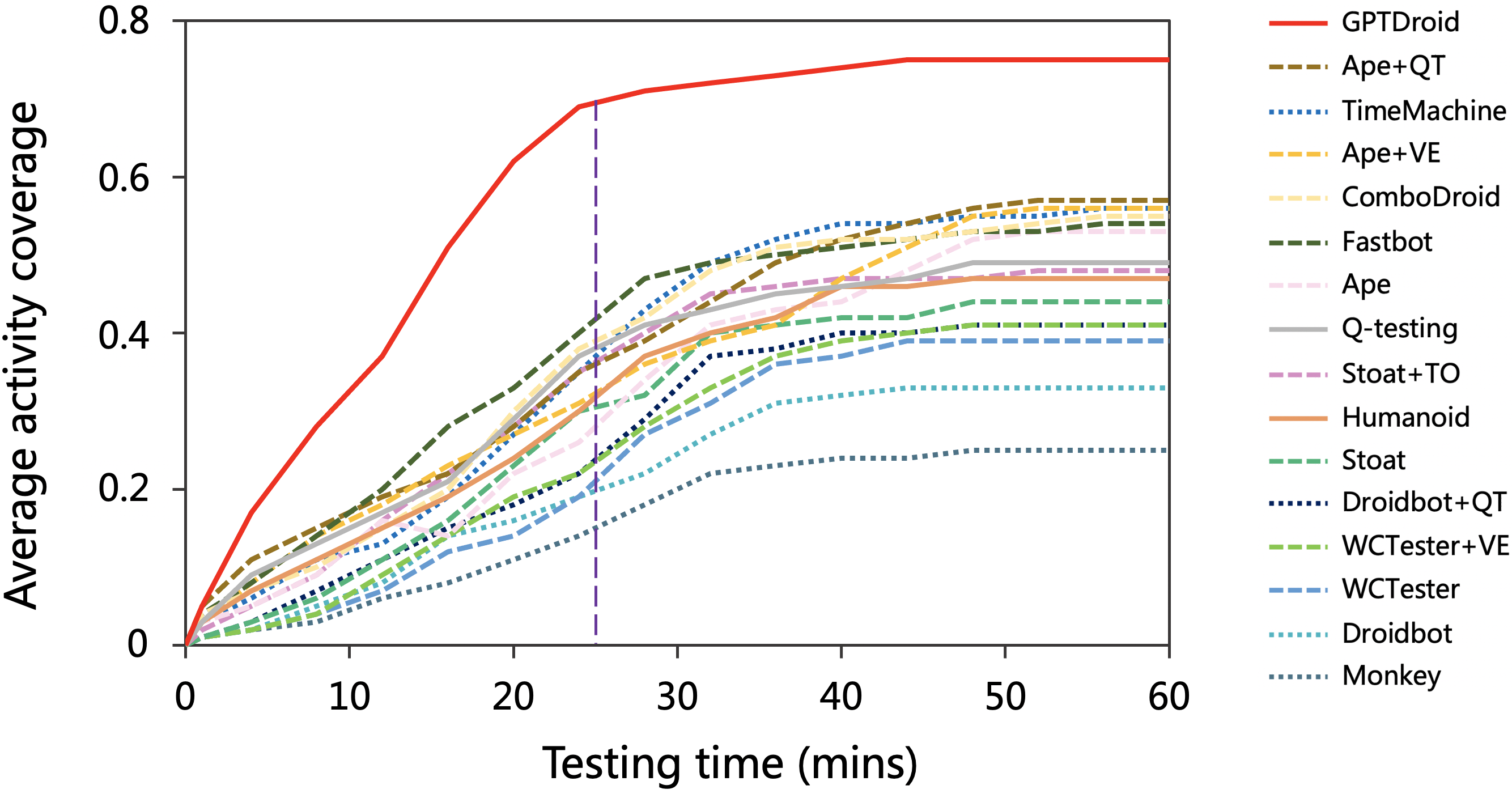}
\vspace{-0.1in}
\caption{Activity coverage with varying time (RQ1).}
% \chen{Since we mention our ICSE'23 work here, do we need to add it as one baseline?}}
% \chen{1. Please add the (mins) to the x-axis label. 2. For the legend, align the order of the line and the tool name, so that the first red line corresponds to our tool name appearing first in the legand. It also applies to other figures.}}
\label{fig:RQ1-ActivityCoverage}
\vspace{-0.05in}
\end{figure}

%% file: figure/RQ2-result.tex
\begin{figure}[htb]
\centering
\vspace{-0.05in}
\includegraphics[width=8.2cm]{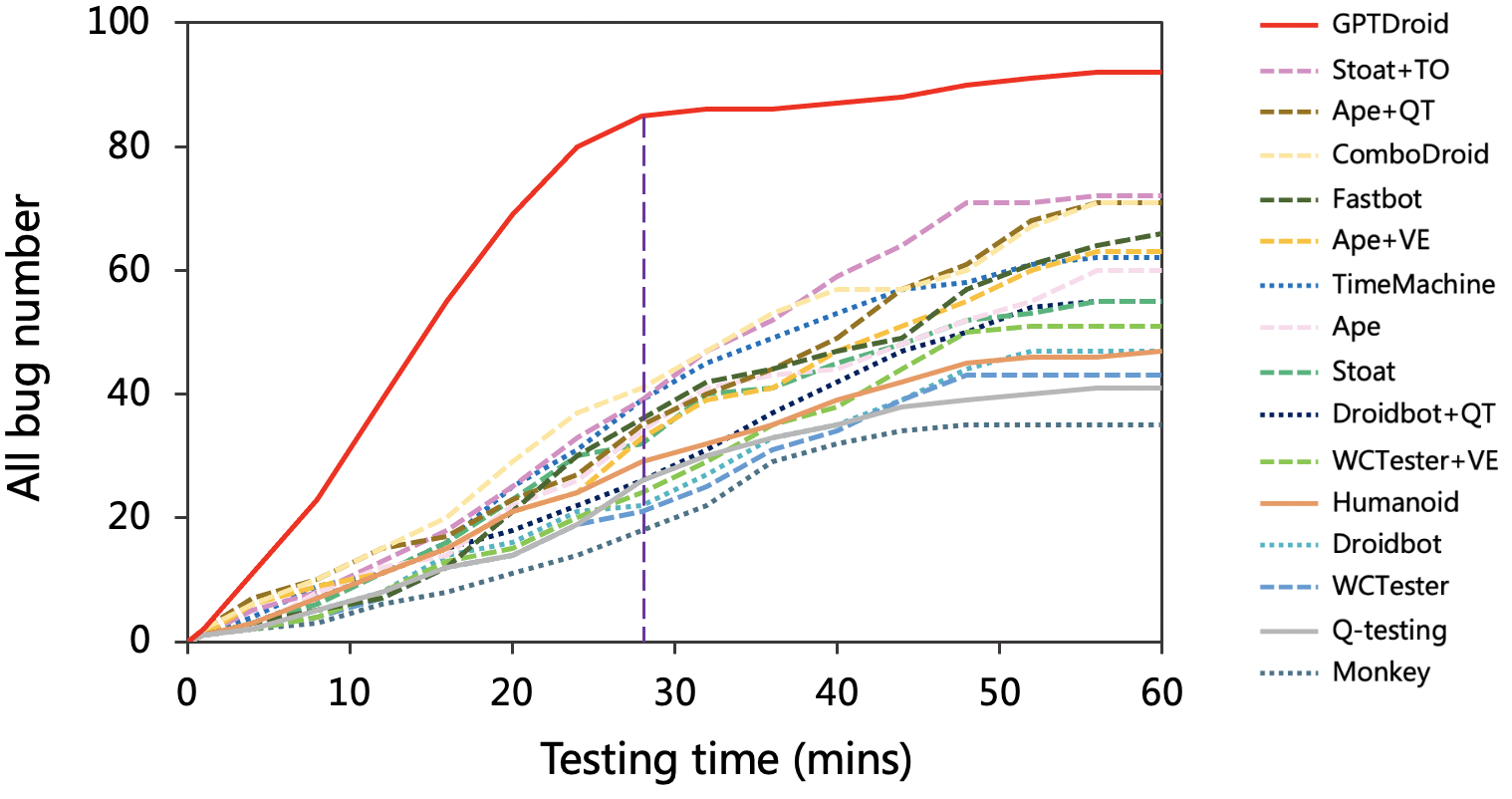}
\vspace{-0.1in}
\caption{Bug detection with varying time (RQ2).}
% \jie{comment}
%这个baselines不应该出现，不是所有的都是baselines。标题也要改改，大小写也不对。就叫做 bug detection performance (RQ2); RQ1那个可能也不能叫做performance，能叫activity coverage 
% }
\label{fig:RQ2-result}
\vspace{-0.1in}
\end{figure}

%% file: tab/RQ3-1-result.tex
\begin{table}[htb]
\vspace{-0.05in}
\renewcommand\arraystretch{1} 
\caption{Contribution of different modules (RQ3)}
\vspace{-0.05in}
\label{tab:RQ3-1-compents}
\centering
\footnotesize
\begin{tabular}{p{2.3cm}<{\centering} |  p{2.2cm}<{\centering} | p{2.0cm}<{\centering}}
% \hline
\toprule
% \multirow{2}*{} & \multicolumn{2}{c}{\textbf{Average coverage}}  \cr 
\textbf{Module} & \textbf{Activity coverage} & \textbf{Code coverage} \\
 \midrule
\textbf{{\tool}} & \textbf{0.75} & \textbf{0.66} \\
\midrule
\textit{w/o GUI Context }  & 0.17 & 0.14 \\
\textit{w/o Function Memory} & 0.34 & 0.28 \\
\bottomrule
\end{tabular}
% \vspace{0.05in}
% \begin{tablenotes}
% % \footnotesize
% \scriptsize
% \item \textbf{\textit{Notes:}} The two variants respectively denote {\tool} removing module 1 (context extraction) and module 3 (operation memorizer). \chen{Why not give the abbreviation for each module, rather than using 1, 2, 3...}
% \end{tablenotes}
\vspace{-0.1in}
\end{table}

% \begin{table}[!t]
% \vspace{-0.05in}
% \renewcommand\arraystretch{1} 
% \caption{Contribution of different modules (RQ3)}
% \vspace{-0.05in}
% \label{tab:RQ3-1-compents}
% \centering
% \footnotesize
% \begin{tabular}{p{2.3cm}<{\centering} | p{1.0cm}<{\centering} | p{1.3cm}<{\centering} || p{1.2cm}<{\centering} | p{1.0cm}<{\centering}}
% % \hline
% \toprule
% \multirow{2}*{\textbf{Module}} & \multicolumn{2}{c||}{\textbf{Number}} & \multicolumn{2}{c}{\textbf{Average coverage}}  \cr 
%  & \textbf{\#Widgets} & \textbf{\#Activities} & \textbf{Activity} & \textbf{Code} \\
%  \midrule
% \textbf{{\tool}} & \textbf{5243}  & \textbf{1049}  & \textbf{0.75} & \textbf{0.66} \\
% \midrule
% \textit{w/o GUI Context } & 573 & 239 & 0.17 & 0.14 \\
% \textit{w/o Function Memory} & 1807 & 477 & 0.34 & 0.28 \\
% \bottomrule
% \end{tabular}
% % \vspace{0.05in}
% % \begin{tablenotes}
% % % \footnotesize
% % \scriptsize
% % \item \textbf{\textit{Notes:}} The two variants respectively denote {\tool} removing module 1 (context extraction) and module 3 (operation memorizer). \chen{Why not give the abbreviation for each module, rather than using 1, 2, 3...}
% % \end{tablenotes}
% \vspace{-0.1in}
% \end{table}

%% file: figure/RQ3-2-result.tex
\begin{figure}[htb]
\centering
\vspace{-0.05in}
\includegraphics[width=7.6cm]{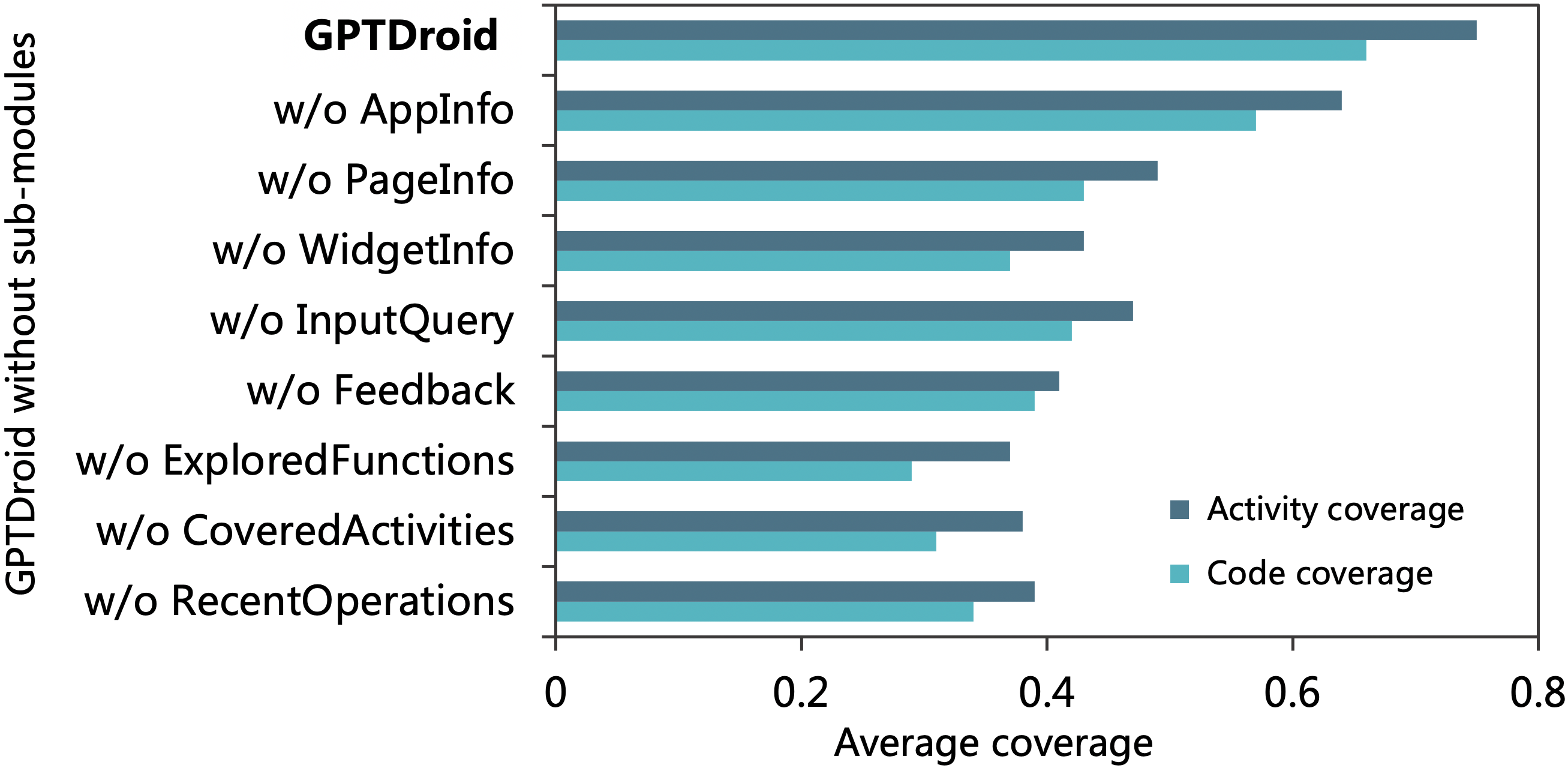}
\vspace{-0.1in}
\caption{Contribution of different sub-modules (RQ3).}
% \jie{comment}}
%感觉gptdroid这个红色有点突兀，要不就用黑色加粗就行？
% \jie{comment}
%这个顺序还需要调整下，input往前，feedback靠后，input那个叫做 inputQuery吧。而且单复数要统一一下，function是单数，activities是复数。
% }
\label{fig:RQ3-2-result}
\vspace{-0.1in}
\end{figure}

%% file: figure/RQ3-3-result.tex
\begin{figure}[htb]
\centering
\vspace{0.05in}
\includegraphics[width=7.6cm]{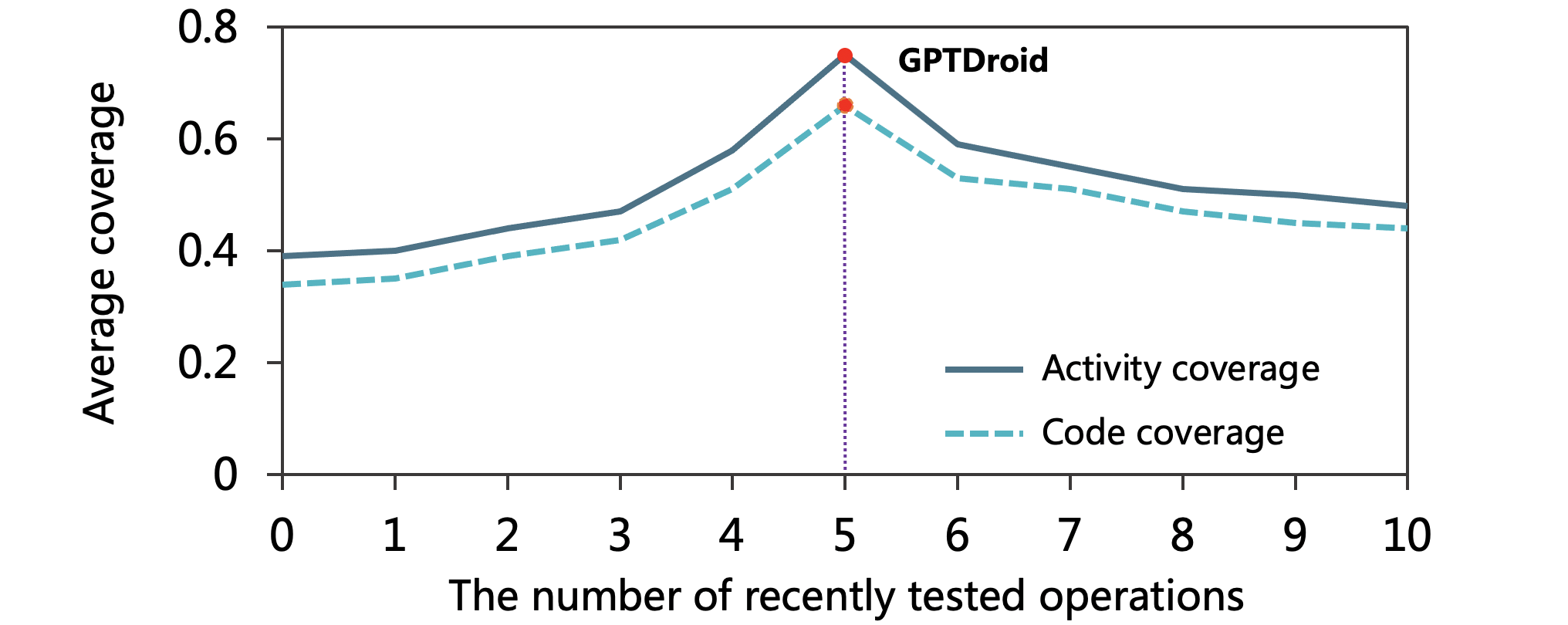}
\vspace{-0.1in}
\caption{Different number of tested operations (RQ3).}
\label{fig:RQ3-3-result}
\vspace{-0.1in}
\end{figure}

%% file: figure/Case-study.tex
\begin{figure*}[t]
\centering
\vspace{0.2in}
\includegraphics[width=17.5cm]{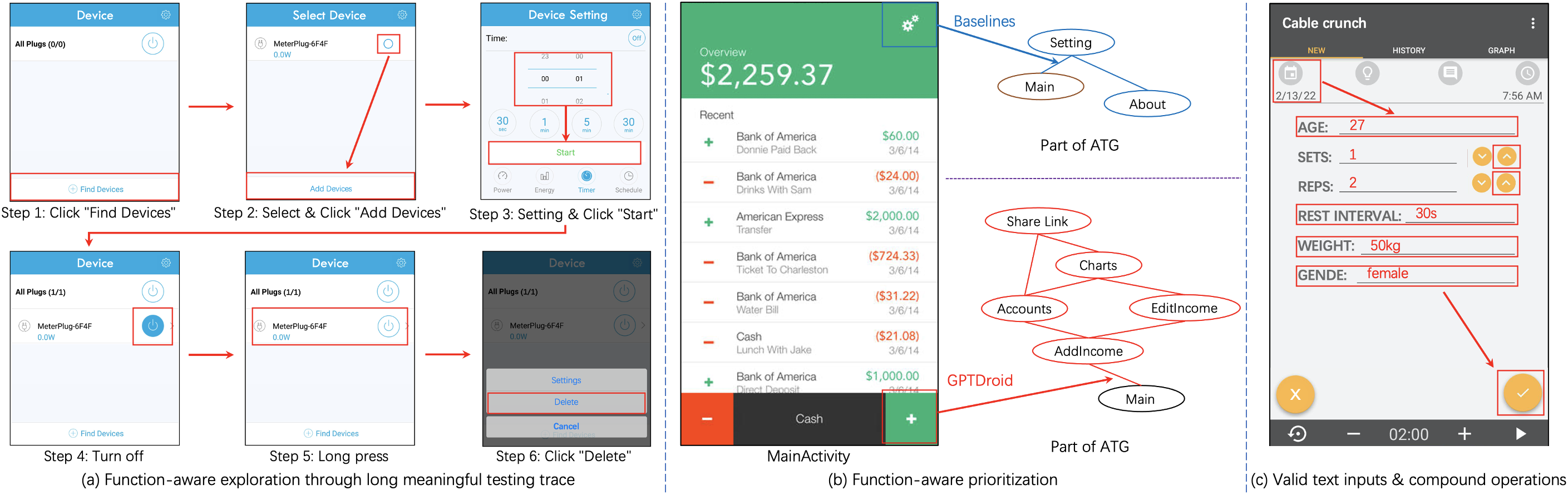}
\vspace{-0.05in}
\caption{Examples of our insights from experiments.}
% \chen{Try to reduce the padding space and make the figure larger. (a) cut the advertisement at the end and make it larger. (b) Does that example really need hardware support i.e., device? Otherwise, we may be attacked.}}
% \jie{comment}
%(a)图里面那个multiple text input可以删掉了。
%b图里面的step尽量简洁，也能更好的看清，例如 step 4 turn off即可。5，也不用出现那个名字了。
% } 
\label{fig:case-study}
% \vspace{-0.05in}
\end{figure*}

%% file: sec/usefulness.tex
\section{Usefulness Evaluation}
\label{sec_Usefulness}

\subsection{Experimental Setup}
\label{sub_Usefulness_Experimental_Setup}

This Section further evaluates the usefulness of {\tool} in detecting new crash bugs.
% To further assess the usefulness of our {\tool} \jie{in detecting new bugs},
We employ a similar experimental setup to the previous section. 
To make it brief, we only compare the best baselines, i.e., TimeMachine, Stoat with Toller and Humanoid, the best one from each type of method following their bug detection performance as shown in Figure \ref{fig:RQ2-result}.
Note that, for coverage, the top three are different sets of baselines, yet this Section is concerned about their capability in detecting bugs, so we use their bug detection performance as the criteria.
% \jie{comment}
%现在这里就有问题。因为我们新增了复合法方法，这些方法已经不是最好的了。所以我们这里要不就不比较基线了，因为之前也已经比较过了，正好也可以稍微缩减下篇幅
% (random-/rule-based methods, model-based methods and learning-based methods)
% Different from it, we choose three baselines, which perform best in bug detection
% \jie{comment}
%这里有个小问题，你选的感觉是activity coverage最好的，但你这里是detect bugs，感觉应该选bug detection最好的，是吧？要不这里改一改吧，我觉得他们应该都发现不了。应该不是timeMachine了，是stoat，其他两个不知道是不是，核实一下。
% in the previous section \chen{Why one, mention specifically.} respectively from three types of methods (random-/rule-based methods, model-based methods and learning-based methods), i.e., Droidbot, Stoat and Humanoid.

We begin with the most popular and recently updated 317 apps from 12 categories as in the previous Section. 
Then we reuse the four criteria in Section \ref{subsec_experiment_dataset} for filtering out the unusable apps.
Differently, we loosen criteria 4, which only requires the app to have ways for bug reporting, since the issue records or pull requests are not mandatory in this Section.
This results in 223 apps for usefulness evaluation. 
% \chen{Where do you get these apps? Different from last section. Still, I got pretty confused with the data collection in this and last section. Please reshape it.}
Note that, this Section aims at evaluating whether {\tool} can detect new bugs in the apps, thus the overlap between the apps of this Section and the previous Section is allowed.
We use the same hardware and software configurations as the previous evaluation Section. 
When the crash bugs are detected, we report them to the development team through online issue reports or emails.

\subsection{\textbf{Results and Analysis}}
\label{subsub_Effectiveness_Result}

For the 223 apps, {\tool} detects 135 bugs involving 115 apps, of which 53 bugs involving 41 apps are newly-detected bugs. 
Furthermore, only 9 of these new bugs were detected by three baselines.
% Furthermore, these new bugs are not detected by the three baselines. 
We submit these 53 bugs to developers, and 35 of them have been fixed/confirmed so far (20 fixed and 15 confirmed), while the remaining are still pending (none of them is rejected). 
This further indicates the effectiveness of our proposed {\tool} in bug detection.
Due to space limit, Table \ref{tab:RQ3-Usefulness} presents these fixed/confirmed bugs, and the full lists can be found on our website.

We further analyze the details of these bugs, and 17 of them involve multiple text inputs or compound operations. 
Besides, we also observe that there are 11 bugs with more than 20 operations before triggering the bug, counting from the \textit{MainActivity} page, which indicates the ability of {\tool} in testing deeper features.
Furthermore, we find at least 28 bugs related to the main business logic of the app.
% , for example, a bug about the health data statistics is revealed for a digital health app.
This further demonstrates the capabilities of {\tool}, and we provide analysis in Section \ref{Sec_Discussion}. 
% This further demonstrates the capabilities of {\tool}, and we provide a more detail analysis in Section \ref{Sec_Discussion}. 
%All these observations again demonstrate the powerful capability of our approach in understanding the semantics of the app, and conducting complex operations and meaningful operation sequences. 

\input{tab/usefulness.tex}

% We further analyze the test path where we found these bugs, and 17 of them have complex operations in the test path. In addition, we also observe that from the \textit{MainActivity} page as a starting point, there are 11 bugs with more than 20 steps to explore, which also shows that {\tool} is good at detecting deeper bugs. We further analyze the relationship between bugs and the main business logic of the application. We find that 28 bugs appeared in the main business logic of the app, such as the digital health app, whose bugs appeared in the health data statistics. 
% \chen{Need to briefly mention that these bugs confirm findings in the last RQ in last section.}

%% file: tab/usefulness.tex
\begin{table}[h]
% \vspace{-0.1in}
\caption{Confirmed or fixed bugs} 
% \chen{Necessary to list the last three columns?}}
%\chen{May put the app package name and links to our GitHub website}}
\vspace{-0.1in}
\label{tab:RQ3-Usefulness}
\centering
% \footnotesize
\scriptsize
\begin{tabular}{p{0.3cm}<{\centering} | p{1.0cm}<{\centering} | p{1.0cm}<{\centering} | p{1.0cm}<{\centering} | p{1.1cm}<{\centering} | p{0.3cm}<{\centering} | p{0.6cm}<{\centering} | p{0.3cm}<{\centering}}
\toprule
\textbf{Id} & \textbf{APP Name} & \textbf{Category} & \textbf{Download} & \textbf{Status} & \textbf{TM} & \textbf{ST+TO} & \textbf{HM} \\
\midrule
1 & PerfectPia & Music & 50M+ & confirmed & & & \\

2 & MusicPlayer & Music & 50M+ & confirmed & & & \\  

3 & NoxSecu & Tool  & 10M+ & fixed & \checkmark & & \\

4 & INSTA & Finance & 10M+ & fixed & & \checkmark & \\

5 & Degoo& Tool & 10M+ & fixed & & & \\  

6 & Proxy & Tool & 10M+ & confirmed & & & \\  

7 & Secure & Tool & 10M+  & fixed & & & \\  

8 & Revolut & Finance & 10M+  & fixed & & & \\  

9 & Thunder & Tool & 10M+  & confirmed  & \checkmark & \checkmark & \\

10 & ApowerMir & Tool & 5M+  & fixed & & & \\  

11 & MediaFire & Product & 5M+  & confirmed & & \checkmark & \\

12 & WAVMoney & Finance & 1M+ & fixed & & & \\

13 & Postegro & Commun & 500K+ & fixed & & & \\   

14 & Deezer MP & Music & 500K+  & fixed & \checkmark & & \\  

15 & MTG & Utilities & 500K+ & fixed  & & & \\  

16 & Yucata & Tool & 500K+ & confirmed  & & & \\  

17 & ClassySha & Tool & 500K+  & fixed  & & & \\  

18 & Linphone & Commun & 500K+  & confirmed  & \checkmark & \checkmark & \\ 

19 & OFF & Health & 500K+ & confirmed  & & & \\ 

20 & Paytm & Finance & 100K+ & confirmed  & & & \\

21 & Transdroid & Tool & 100K+ & confirmed  & & \checkmark & \checkmark \\ 

22 & Transistor & Music  & 10K+ & fixed  & & & \\

23 & Onkyo & Music  & 10K+  & fixed  & & & \checkmark \\

24 & Democracy & News  & 10K+ & confirmed  & & & \\

25 & NewPipe  & Media  & 10K+ & confirmed  & \checkmark & & \\ 

26 & LessPass & Product  & 10K+ & confirmed  & & & \\  

27 & CEToolbox &  Medical  & 10K+  & confirmed  & & & \\ 

28 & OSM  & Health  & 10K+  & fixed  & & & \\  

29 & Monse  & Finance  & 10K+  & fixed  & & & \checkmark \\  

30 & Fitb  & Health  & 10K+  & confirmed  & & &  \\  

31 & KHAN  & Education  & 10K+  & fixed  & & &  \\  

32 & Leaprt  & Tool  & 10K+  & fixed  & & &  \\  

33 & Penly  & Product  & 10K+  & fixed  & & &  \\  

34 & Rocket  & Product  & 10K+  & fixed  & & &  \\  

35 & Fkowy  & Tool  & 10K+  & fixed  & & &  \\  
\bottomrule

\hline
\end{tabular}
% \vspace{-0.15in}
\end{table}

%% file: sec/discussion.tex
\section{Insights from Experiment Results}
\label{Sec_Discussion}

This Section summarizes 4 kinds of capabilities of {\tool} including high-level (i.e., long meaningful test trace, test case prioritization) and low-level ones (i.e., valid text input, compound actions), to unveil the mystery of why {\tool} outperforms existing method.  

% \rev{Despite the superior performance of {\tool} in the last section, it is still unclear the reason behind it.
% To fully understand the testing capability of the LLM, we carry out a qualitative study by investigating the cases in which our model outperforms baselines.
% We summarize four kinds of capabilities including low-level (i.e., valid text input, and compound actions) and high-level ones (i.e., long meaningful test trace, and test case prioritization). 
% These findings pave the way for further research in this area. }

% \textbf{Long meaningful test trace.}
\textbf{Functionality-aware exploration through the long meaningful testing trace.}
{\tool} can automatically generate the test cases with a long sequence of operations which together accomplish a business logic of the app, and this is quite important for covering the app features and ensuring its quality. 
As shown in Figure \ref{fig:case-study} (a), in SmartMeter app~\cite{SmartMeter}, to test a commonly-used app feature ``delete equipment'', the automated tool first needs to click ``Find Devices'' on the device page, then select a device (Bluetooth is turned on and there are candidate devices) and click ``Add Devices'' for adding it in the device page, input the related information and click ``Start'' to start the device, then turn off this device in the device page, long press it and click ``Delete'' from the pop-up menu.
Only with this long sequence of operations that touches the ``deleting equipment'' feature, a crash can be revealed.
Our functionality-aware memory prompt can enable the LLM to capture the long-term dependencies among GUI pages to conduct the functionality-guided exploration. 
% \rev{It may be because of ChatGPT's exposure to tutorial or bug reports which contain step-by-step instructions or descriptions of how to trigger a certain feature or reproduce a certain bug in the training corpus.}\cite{su2021benchmarking,wang2022detecting} 
% \rev{ At the same time, the Long-term and Short-term memory we provide will help ChatGPT understand the test process and make reasonable judgments through the analysis of the test sequence.
% Therefore, provided with low-level semantic information (i.e., current GUI page) and high-level testing history, {\tool} can capture long-term dependencies among GUI pages for generating long meaningful exploration sequences.}

% \textbf{Important widget prioritization.} 
\textbf{Function-aware prioritization.} 
We also observe that {\tool} usually prioritizes testing the ``important'' functions, which is valuable for reaching a higher activity coverage and covering more key activities with relatively less time.
As shown in Figure \ref{fig:case-study} (b), on the Main page of the Moni app~\cite{Moni}, the baseline tools tend to first click the ``Setting'' button following the exploration order from upper to lower, which leads the testing easily trapped into the setting page cycle.
{\tool} chooses to first click the ``AddIncome'' button to explore the add income functionality which is the key feature of the app. 
This is facilitated by the semantic understanding of the GUI page and the functionality-aware memory designed in {\tool}.

% test the ``AddIncome'' button, i.e., click the ``add'' button, which is based on the semantics of the GUI page and app information, and can quickly explore the activities related to the key features of the app. 

% We also observe that {\tool} usually prioritizes testing the ``important'' widgets, which is valuable for reaching a higher activity coverage and covering more key activities with relatively less time.
% As shown in Figure \ref{fig:case-study} (b), in the Main page of the Moni app~\cite{Moni}, the baseline tools tend to first click the ``Setting'' button following the exploration order from upper to lower, which leads the testing easily trapped into the setting page cycle.
% {\tool} chooses to first test the ``AddIncome'' activity, i.e., click the ``add'' button, which is based on the semantics of the GUI page and app information, and can quickly explore the activities related to the key features of the app. 
% This 
% \rev{This may be because that ChatGPT's training corpus contains a wide variety of software-related information, including user manuals, release notes, and app descriptions where developers often highlight the most important features at the front.
% At the same time, we have also added the current application testing function to Long-term memory, which helps GPTDroid test important functions first in combination with functions.}

% \jie{comment}
%这里我觉得应该调整下顺序。把3放最前面，然后是4，应该突出整体的作用。3的名字可以叫做funtion-aware exploration through long meaningful test trace。得突出测试的是功能。
%然后4，是不是也可以改成funtion-aware prioritization ？ 这个我不确定。讲真我没有特别看到这个表现和我们方法设计之间的关系。

\textbf{Valid text inputs.} {\tool} can automatically fill in valid text content to the input widget which is essentially the key for passing the page as seen in Figure \ref{fig:case-study} (c).
Similar to prior work~\cite{liu2022fill}, our model can generate semantic text input (e.g., income, date, etc) accordingly.
Besides single text input, it can also successfully fill in multiple input widgets at the same time which are correlated to each other like the departure and arrival cities and dates in the flight booking app.
We have designed a prompt, especially for querying the text input and utilize the few-shot learning by providing demonstrations with the output template to facilitate the LLM generating desired executive commands, which can be accurately mapped to the GUI widgets of text inputs to enable it to execute automatically. 
% \rev{It may be due to ChatGPT's language generation capabilities which was well learned during the training phase. 
% Further validated the effectiveness of our prompt designed for text input in Section }
% \ref{subsec_approach_information_Extraction}.
% ~\cite{schulman2022chatgpt}

\textbf{Compound actions.} {\tool} can conduct complex compound operations guided by the LLM. 
As shown in (Figure \ref{fig:case-study} (c), to add the ``Cable crunch'' information, it first inputs the text, selects the date, sets the ``SETS'' and ``REPS'' by clicking the upper or lower button, then click the submit button in the lower right corner. 
Thanks to our designed executive command generation method with few-shot learning and output template, {\tool} can accurately map the LLM's output into the actions related to GUI widgets.

%% file: sec/related.tex
\section{Related Work}
\label{sec_related}

\textbf{Automated GUI testing.} To ensure the quality of mobile apps, many researchers study the automatic generation of large-scale test scripts to test apps~\cite{xie2007designing}.
% In Android, test script generators interact with apps in the same way as humans: sending simulated gestures to the GUI of an app. Since the acceptable gestures in a UI state are limited, the main difference between different test generators is their strategies used in prioritizing these actions.
Monkey~\cite{Monkey} is the popular random-based automated GUI testing tool, which emits pseudo-random streams of UI events and some system events. 
% Monkey is widely used in industry for stress testing because 
% It is easy to use and compatible with different Android versions.  
However, the random-based testing strategy cannot formulate a reasonable testing path according to the characteristics of the app, resulting in low test coverage. 
To improve the test coverage, researchers propose model-based~\cite{mirzaei2016reducing,yang2018static,yang2013grey,zeng2016automated,mao2016sapienz,su2017guided,dong2020time,gu2019practical,wang2020combodroid} automated GUI testing methods, design corresponding models through the research and analysis of large-scale apps.
% , and analyze the app code in real-time during the test process to formulate reasonable test strategies.
% Sapienz~\cite{mao2016sapienz} used genetic algorithms as the model and Stoat~\cite{su2017guided} used the stochastic model learned from an app to optimize test suite generation. 
% Ape~\cite{gu2019practical} used the runtime information to dynamically evolve its abstraction criterion via a decision tree and generated UI events via a random and greedy depth-first state exploration strategy.
% ComboDroid~\cite{wang2020combodroid} obtained such use cases either from humans or automatically generates from a GUI model constructed by GUI exploration and analyzed the data flow between obtained use cases, and combined them to generate final tests. 
Although model-based automated GUI testing tools can improve test coverage, the coverage is still low because it does not consider the semantic information of the app's GUI and Page.
Researchers further proposed human-like testing strategies and designed learning-based~\cite{li2019humanoid,pan2020reinforcement} automated GUI testing methods. 
% Humanoid~\cite{li2019humanoid} used a deep neural network model that predicts which UI elements on the current GUI page are more likely to be interacted with by users and how to interact with them.
% Q-testing~\cite{pan2020reinforcement} used a reinforcement learning-based method to compare GUI pages and give rewards. These rewards are used and iteratively updated to guide the testing to cover more functionalities of apps.
Although the learning-based approach can improve the test coverage by learning the interactive processes or using the idea of reinforcement learning. However, it is still unable to better understand the semantic information of the page and plan the path according to the actual situation of the app.
We aim at proposing a more effective approach to generate human-like actions for thoroughly and more effectively testing the app, accomplishing it with LLM. There are also studies that tried to find bugs in similar apps to move beyond crashes \cite{tan2020collaborative}, yet it cannot reveal the crashes automatically as this work. 
Techniques related to test migration \cite{Behrang2019test,Talebipour2021uitest} can generate meaningful operation sequences borrowed from the source app, yet it is quite demanding to require the test cases of an app, and by comparison, {\tool} can generate more meaningful test traces from scratch.

% focuses on a different direction for improving the activity coverage, i.e., encoding the visual information of the application GUI into natural language description for prompt learning, and conducting operations following the LLM's feedback.
% real-time planning is carried out according to the actual situation of the page in the test process.

% \subsection{LLM for Software Engineering}
\textbf{LLM for Software Engineering.}
% Recently, there has been a great success of pre-trained Large Language Models (e.g.,  RoBERTa~\cite{2019RoBERTa}, GPT-3~\cite{brown2020GPT3}, PaLM~\cite{chowdhery2022palm},
% OPT~\cite{zhang2022opt}) in a variety of NLP tasks. 
% Due to the large amounts of available pre-training data from the internet, research shows that LLMs can already be used for very specific downstream tasks through the new paradigm ``pre-train, prompt and prediction''~\cite{liu2023pre} without any fine-tuning of special data sets.
% This paradigm for LLM was widely used in many work and achieved state-of-the-art performance on downstream tasks~\cite{huang2022prompt,deng2022fuzzing}. The core of this paradigm is to use prompt engineering~\cite{liu2023pre,Branwen2020Gpt-3creative, Cantino201Prompt,ge2021visual}, where a natural language description of the task is provided to the LLM. 
Considering the powerful performance of LLM, researchers have successfully leveraged it to solve various tasks in the field of software engineering~\cite{23a3testTestGeneration,25adaptiveTestGeneration,63chatUniTest,66generatingUnitTests,64noUnitTest,39assertStatements,24semanticsTestCompletion,56retrievalPromptSelection,79StudyingtheUsage,80UsingTransferLearning,14conformanceTesting,1itigerIssueTitle,54explainableDebugging,40circleContinualRepair,35vulnerabilityRepair}.
Supported by code naturalness~\cite{hindle2016naturalness}, researchers applied the LLMs to code writing in different programming languages~\cite{chen2021evaluating,feng2020codebert,fried2022incoder,xu2022systematic}. 
% Huang et al.~\cite{huang2022prompt} used LLM for the type inference in statically-typed partial Code. 
In testing, LLMFuzz~\cite{deng2022fuzzing} used LLMs to generate input programs for fuzzing Deep Learning libraries. 
Xia et al.~\cite{xia2022less} applied LLM for automatic program repair to improve the accuracy of the generated repair patches. 
Lemieux et al. \cite{lemieux2022codamosa} leveraged LLM in escaping the coverage plateaus in test generation.
Kang et al. \cite{kang2023large} explored the LLM-based bug reproduction. 
A similar work QTypist \cite{liu2022fill} leveraged the LLM to generate the text inputs for passing a GUI page in order to improve the testing coverage of mobile testing.
Different from its sole focus on text input generation to boost existing GUI testing tools, {\tool} is a complete GUI testing tool in asking LLM to propose different actions to interact with the target app including clicking buttons, filling in text and even more complicated compound actions.
It makes this work more generalized for wild mobile app testing.

%% file: sec/conclusion.tex
\section{Conclusion}
\label{sec_conclusion}

Automated GUI testing has made much progress, yet still suffers from low activity coverage and may miss critical bugs.
This paper aims at generating human-like actions to facilitate app testing more thoroughly and effectively. 
Inspired by ChatGPT, we formulate the GUI testing problem as a Q\&A task and propose {\tool}.
It extracts the GUI context and functionality-aware memory, encodes them into prompt questions to ask the LLM, decodes the LLM's feedback answer into actionable operations to execute the app, and iterates the whole process. 
Results on 93 popular apps demonstrate that {\tool} can achieve 75\% activity coverage, with 32\% higher than the best baseline, and can detect 31\% more bugs with faster speed than the best baseline. 
{\tool} also detects 53 new bugs on Google Play with 35 of them being confirmed/fixed.
% The capability of {\tool} in generating semantic text input and compound actions, guiding to explore the long meaningful test trace, and prioritizing test cases, further proves the effectiveness and human-like aspects of {\tool}.
In the future, we plan to explore more advanced prompt engineering design to better exploit the power of LLM, and may also fine-tune open-source LLM for this specific tasks for better performance.